\documentclass[useAMS,usenatbib]{mn2e}
\usepackage[dvips]{graphicx}
\usepackage{amsmath,amssymb}
\usepackage{color}

\title[Gas velocity patterns as a test of spiral theories]
{Gas velocity patterns in simulated galaxies:
Observational diagnostics of spiral structure theories}
\author[J. Baba et al.]
{J. \textsc{Baba}$^{1,2}$\thanks{E-mail:babajn@elsi.jp; baba@cosmos.phys.sci.ehime-u.ac.jp}, 
K. Morokuma-Matsui$^{3,4}$,
Y. Miyamoto$^{3}$,
F. Egusa$^{4}$
and N. Kuno$^{3,5}$
\\ 
$^{1}$Earth-Life Science Institute, Tokyo Institute of Technology, 2-12-1 Ookayama, Meguro, Tokyo 152--8551, Japan.\\
$^{2}$Research Center for Space and Cosmic Evolution, Ehime University, Bunkyo-cho 2-5, Matsuyama 790-8577, Japan\\
$^{3}$Nobeyama Radio Observatory, National Astronomical Observatory of Japan, 462-2 
Nobeyama, Minamimaki, Minamisaku, Nagano 384-1305, Japan\\
$^{4}$Chile Observatory, National Astronomical Observatory of Japan, 2-21-1 Osawa, Mitaka, Tokyo 181--8588, Japan\\
$^{5}$Division of Physics, Faculty of Pure and Applied Sciences, University of Tsukuba, Ten-noudai, Tsukuba, Ibaraki 305-8571, Japan
}

\begin{document}

\date{Accepted 2016 April 22. Received 2016 April 22; in original form 2016 February 12}

%\pagerange{\pageref{firstpage}--\pageref{lastpage}} \pubyear{2002}

\maketitle

\begin{abstract}
There are two theories of stellar spiral arms in isolated disc galaxies that model stellar spiral arms with different longevities:
quasi-stationary density wave theory, which characterises spirals as rigidly rotating, long-lived patterns (i.e. steady spirals), 
and dynamic spiral theory, which characterises spirals as differentially rotating, transient, recurrent patterns (i.e. dynamic spirals). 
In order to discriminate between these two spiral models observationally, we investigated the differences 
between the gas velocity patterns predicted by these two spiral models in hydrodynamic simulations.
We found that the azimuthal phases of the velocity patterns relative to the gas density peaks (i.e. gaseous arms) 
differ between the two models, as do the gas flows; nevertheless, the velocity patterns themselves are similar for both models. 
Such similarity suggests that the mere existence of streaming motions does not conclusively confirm the steady spiral model.
However, we found that the steady spiral model shows that 
the gaseous arms have {\it radial} streaming motions well inside the co-rotation radius, 
whereas the dynamic spiral model predicts that the gaseous arms tend to have {\it tangential} streaming motions. 
These differences suggest that the gas velocity patterns around spiral arms will enable distinction between the spiral theories. 
\end{abstract}


%%%%%%%%%%%%%%%%
\section{Introduction}
\label{sec:intro}

Spiral structures are the most prominent features in disc galaxies.
Near-infrared observations have shown that spiral structures are gravitationally driven variations in the surface densities 
of stellar discs \citep[][]{Block+1994,RixZaritsky1995,Grosbol+2004,Elmegreen+2011},
which strongly suggests that spiral arms originate from the stellar dynamics in disc galaxies.
Furthermore, observations of nearby spiral galaxies have shown that 
most of the star formation in spiral galaxies is associated with spiral arms 
\citep[][]{LordYoung1990,CepaBeckman1990,SeigarJames2002,GrosbolDottori2009}. 
Accordingly, understanding how spiral arms in galaxies form and evolve is essential to
understand both galactic dynamics and star formation, i.e. galaxy evolution.
In this paper, we focus on spiral arms that arise in disc galaxies without external perturbations.

Major progress in theoretical studies on spiral structures was made in the 1960s--1980s by hypothesizing
stellar spiral arms (hereafter, spiral arms) to be quasi-stationary density waves 
\citep[][for a review]{Lindblad1963,LinShu1964,LinShu1966,Lin+1969,Bertin+1989a,Bertin+1989b,BertinLin1996}.
In this so-called `quasi-stationary spiral structure' (QSSS) hypothesis, spiral arms are supposed to be 
rigidly rotating, long-lived patterns (hereafter, `steady spiral') that persist for 
at least several galactic rotations (i.e., $\gtrsim$ 1 Gyr). Consequently, these steady spirals affect gas flows;
when the gas inside a co-rotation radius ($R_{\rm CR}$) {\it overtakes} a spiral arm as it moves around a galactic disc, 
the gas is expected to form a standing shock, called a `galactic shock', around the spiral arm
\citep[][]{Fujimoto1968,Roberts1969,Shu+1972,Shu+1973,Sawa1977,IshibashiYoshii1984,LeeShu2012,Lee2014} 
within one or two transits of the gas through the spiral arm 
\citep[$\lesssim 100$ Myr;][]{Woodward1975,RobertsHausman1984,WadaKoda2004,Wada2008}.
The QSSS/galactic shock hypothesis predicts that the gas velocity changes suddenly; 
it is believed that such velocity changes are observed as wiggles in isovelocity contours of $\rm H_I$/CO gas 
around the spiral arms of nearby galaxies such as M81 \citep[e.g.][]{RotsShane1975,Rots1975,Visser1980b} and 
M51 \citep[e.g.][]{Tully1974a,Rydbeck+1985,Vogel+1988,Rots+1990,
Garcia-Burillo+1993,KunoNakai1997,Aalto+1999,Miyamoto+2014}\footnote{
It is believed that the grand-design spirals of M51 and M81 are driven by tidal interactions with companion galaxies 
\citep[e.g.][]{Toomre1981,Sundelius+1987,HowardByrd1990,ThomassonDonner1993,
SaloLaurikainen2000a,SaloLaurikainen2000b,Dobbs+2010}.
See also Section \ref{sec:discussion}.
}.
Recent multi-dimensional hydrodynamic simulations have predicted 
that the shock (i.e. gaseous spiral arm) locations move {\it monotonically} 
from downstream to upstream of the stellar spiral arm with increasing radius 
\citep{GittinsClarke2004,Martinez-Garcia+2009b,KimKim2014,Baba+2015a}.

On the other hand, \citet{GoldreichLynden-Bell1965} and \citet{JulianToomre1966} proposed 
a transient spiral hypothesis based on linear local analyses of gaseous and stellar discs, respectively.
Numerical simulations of galactic discs have extended this hypothesis 
to a transient recurrent spiral (hereafter, `dynamic spiral') hypothesis \citep[see a review by][]{DobbsBaba2014}.
According to this hypothesis, the amplitudes of spiral arms change on the time-scale of 
a galactic rotation or even less (i.e. a few hundreds of Myrs) for multiple-arm spirals \citep{SellwoodCarlberg1984,
Fujii+2011,Wada+2011,Grand+2012a,Baba+2013,D'Onghia+2013,Pettitt+2015,KumamotoNoguchi2016}, 
unbarred grand-design spirals \citep{Sellwood2011}\footnote{
The longevity of (unbarred) grand-design spirals is a standing problem. 
Some early reports stated that numerical simulations had succeeded in 
reproducing unbarred long-lived grand-design spirals \citep{Thomasson+1990,Zhang1996}.
However, \citet{Sellwood2011} tested the same models and found that two-armed spirals are not single long-lived patterns, 
but rather the superpositions of three or more patterns that each grow and decay.
More recently, \citet{SellwoodCarlberg2014} reported the existence of longer-lived modes,
which survive multiple rotations \citep[see also][]{Roskar+2012,Minchev+2012}.
Nevertheless, \citet{SellwoodCarlberg2014} argued that their results were inconsistent with the idea that 
spiral arms are quasi-stationary density waves because the arms in their simulations changed with time.
The origins of (unbarred) grand-design spirals are unclear, which are beyond of the scope of this study, 
although it is worth mentioning that tidal interactions could create and re-energize unbarred grand-design spirals 
\citep[e.g.][]{ByrdHoward1992,Oh+2015,Pettitt+2016}.
}, 
and barred spirals \citep{Baba+2009,Grand+2012b,Roca-Fabrega+2013,Baba2015c}.
In contrast to the QSSS/galactic shock hypothesis, the dynamic spiral model predicts that 
the gas does not flow through a spiral arm, but rather effectively falls into 
the spiral potential minimum from {\it both sides} of the arm 
\citep[referred to as `large-scale colliding flows';][]{DobbsBonnell2008,Wada+2011}.
Furthermore, the dynamic spiral model shows {\it no systematic} offset between 
the gas density peak location and the spiral arm \citep{Baba+2015a}.

These recent advances in spiral theory require the determination of observational indicators 
that can be used to distinguish between steady and dynamic spirals. 
Some observational tests that could be employed for this purpose
have been proposed \citep[see a review by][]{DobbsBaba2014}.
It was proposed that the existence of radial metallicity distribution breaks 
could be used to test spiral longevity \citep[][]{Lepine+2011,ScaranoLepine2013}.
On the other hand, focusing on the azimuthal direction, \citet{DobbsPringle2010} proposed that 
observations of the azimuthal distributions of age-dated stellar clusters could be used to 
discriminate among the spiral models \citep[see also][]{Wada+2011,Grand+2012b,Dobbs+2014}.
The measurements of the azimuthal colour gradients across spiral arms over 
wide radial ranges could also be employed to test spiral longevity 
\citep{Martinez-GarciaGonzalez-Lopezlira2013,Martinez-GarciaPuerari2014}.
Furthermore, \citet{Baba+2015a} focused on spatial distributions of gas and old stars
and suggested that radial profiles of the gas-star offset angles can be used to distinguish between the two spiral models.
For the {\it kinematics} of gas and stars, \citet{Kawata+2014} performed analyses using snapshots of 
N-body/hydrodynamic simulations of a Milky Way-sized barred spiral galaxy
and predicted observational signatures of dynamic spirals \citep[see also][]{Baba+2009}.
Recently, \citet{Grand+2015} investigated the differences among the peculiar velocity power spectra of stars 
in the Milky Way galaxy predicted by the two spiral models and discussed a possible application to observations of 
gas velocity fields in external galaxies.

In this study, in order to develop a new method of distinguishing between the two spiral models,
we focused on the differences between their gas velocity pattern predictions around spiral arms,
because gas velocity data are more easily obtainable than measurements of stars in external galaxies.
The remainder of this paper is organized as follows:
In Section \ref{sec:models}, we present the models and methods. 
Section \ref{sec:results} describes the differences between the gas velocity patterns predicted by the two spiral models. 
Section \ref{sec:discussion} summarizes our results and 
discusses an application of the results to observational data from spiral galaxies.
We emphasize that this `velocity pattern method' is not the only means of distinguishing between the spiral models. 
In order to determine the nature of spiral galaxies, other methods, 
such as CO-H$\alpha$ offset measurement \citep{Egusa+2009} and gas-star offset measurement \citep{Baba+2015a}, 
are complementary to the velocity pattern method proposed in this paper.

%%%%%%%%%%%%%%%%
\section{Numerical Simulations}
\label{sec:models}

\subsection{Numerical methods}

To investigate the differences between the gas velocity patterns around the spiral arms 
that are predicted by the steady and dynamic spiral models, 
we conducted hydrodynamic simulations of rigidly rotating spiral potentials (i.e. the  steady spiral model; Section \ref{sec:method:steady}) 
and $N$-body/hydrodynamic simulations of stellar and gaseous discs (i.e. the dynamic spiral model; Section \ref{sec:method:dynamic}).
These simulations were performed with an $N$-body/smoothed particle hydrodynamics (SPH) simulation code, 
{\tt ASURA-2} \citep{SaitohMakino2009,SaitohMakino2010}. 
The self-gravity was calculated with the Tree/GRavity PipE (GRAPE) method using
a software emulator of GRAPE known as Phantom-GRAPE \citep{Tanikawa+2013}.

The simulations also took into account radiative cooling and heating due to interstellar far-ultraviolet radiation \citep[FUV;][]{Wolfire+1995}.
The radiative cooling of the gases was determined by assuming an optically thin cooling function, 
$\Lambda(T, f_{\rm H2} ,G_{\rm 0})$, based on a radiative transfer model of photo-dissociation regions 
across a wide temperature ($T$) range of $20~{\rm K} < T < 10^8~{\rm K}$ \citep{Wada+2009}.
Here, the molecular hydrogen fraction, $f_{\rm H2}$, follows the fitting formula given by \citet{GnedinKravtsov2011},
and $G_0$ is the FUV intensity normalized to the solar neighbourhood value.
The normalized FUV intensity of an SPH particle is given by $G_{\rm 0} = G_{\rm 0,thin} e^{-\sigma_{1000} N_{\rm H}}$,
where $N_{\rm H}$ is the total column density of hydrogen, 
$\sigma_{1000} = 2 \times 10^{-21}~\rm cm^2$ is the effective cross-section for dust extinction 
at $\lambda = 1000~\rm \AA$ \citep{DraineBertoldi1996,GloverMacLow2007a}, 
and $G_{\rm 0,thin}$ is the normalized FUV intensity in the optically-thin limit, 
which is given by summing over all stellar particles \citep{GerritsenIcke1997,Pelupessy+2006}.
We determined the time-dependent FUV luminosities of the stellar particles by mapping their ages 
using the stellar population synthesis modelling software PEGASE \citep{FiocRocca-Volmerange1997}.  
$N_{\rm H}$ was computed using a Sobolev-like approximation \citep{Gnedin+2009}.

We implemented sub-grid models for star formation and stellar feedback.
Star formation was incorporated into the simulation as follows.
If an SPH particle (with density $n$, temperature $T$, and velocity $\mathbf{v}$) satisfied the following criteria: 
(1) $n > 100~{\rm cm^{-3}}$; (2) $T < 100~{\rm K}$; and (3) $\nabla\cdot\mathbf{v}<0$; 
then the SPH particle created star particles in a probabilistic manner following the Schmidt law, 
with a local dimensionless star formation efficiency of $C_{\ast}=0.033$ \citep{Saitoh+2008}.
Feedback from type-II supernovae was implemented as thermal energy \citep{Saitoh+2008,SaitohMakino2009},
and ${\rm H_{II}}$-region feedback was considered using a Stromgren volume approach, 
in which the gases around young stars extending out to a radius that is sufficiently large to achieve ionization balance 
were simply defined as having a temperature of $10^4$ K \citep{Baba+2016b}.

\subsection{Dynamic spiral model}
\label{sec:method:dynamic}

In order to investigate the velocity patterns in the dynamic spiral model, 
we used a 3D $N$-body/SPH simulation of a barred spiral galaxy \citep[see][for details]{Baba2015c}.
In this model, the initial axisymmetric model is comprised of stellar and gaseous discs, a classical bulge, and a dark matter halo,
which follow exponential, Hernquist \citep{Hernquist1990}, and Navarro-Frenk-White \citep{Navarro+1997} profiles, respectively.
The dark matter halo is treated as a fixed external potential. 
In this study, the initial numbers of stars and SPH particles were 6.4 and 4.5 million, respectively,
and the gravitational softening length was 10 pc.
Hereafter, we refer to this barred spiral galaxy model as the `DYNAMIC' model.

The DYNAMIC model satisfies the criteria for bar instability \citep{Efstathiou+1982} and 
produces spontaneous stellar bar development.
\citet{Baba2015c} found that the short-term behaviours of spiral arms 
in the outer regions ($R>$ 1.5--2 bar radii; i.e. $R \gtrsim 5$ kpc) can be explained 
by swing amplification theory \citep[e.g.][]{Toomre1981}; consequently, the effects of bars are not negligible 
in the inner regions ($R<$ 1.5--2 bar radii;  i.e. $R \lesssim 5$ kpc). 
Thus, for the DYNAMIC model, we focus on the spiral arms regions with $R > 6$ kpc.

\subsection{Steady spiral model}
\label{sec:method:steady}

We performed hydrodynamic simulations in a static axisymmetric potential, $\Phi_{\rm 0}(R,z)$, 
with a rigidly rotating spiral potential, $\Phi_{\rm sp}(R,\phi,z;t)$.
Hereafter, we refer to this model as the `STEADY' model.
The static axisymmetric potential was produced by using the cloud-in-cell (CIC) mass-assignment scheme 
from the initial condition of the DYNAMIC model.
The initial conditions for the gaseous disc were identical to those of the DYNAMIC model.
The initial number of SPH particles and the gravitational softening length were 
the same as those in the DYNAMIC model.

The gravitational potential of the spiral arm\footnote{
In the present study, we adopted a simple cosine function for the spiral potential, which is shown in eq.(1).
Nevertheless, as argued by \citet{Kalnajs1973}, the potential perturbations from (stellar) spiral arms 
should result from the crowding of `stellar' orbits in a galactic disc.
In fact, more detailed studies of self-consistent steady spiral models have shown that self-consistent spiral potentials are 
not described by simple cosine functions in the azimuthal direction \citep[e.g.][]{Pichardo+2003,Junqueira+2013}.
However, whether the spiral potential is modeled as cosine or Gaussian does not affect our objectives, 
since the both spiral-potential models are symmetric with respect to the spiral.
} was calculated using 
\begin{eqnarray}
&& \Phi_{\rm sp}(R,\phi,z;t) = A(R,z) \frac{z_{\rm 0}}{\sqrt{z^2+z_{\rm 0}^2}} \nonumber \\
&& \times
\cos \left[m \left\{\phi - \Omega_{\rm p}t + \cot i_{\rm sp} \left(\ln\frac{R}{R_{\rm 0}}\right) \right\} \right],
\end{eqnarray}
where $A$, $m$, $i_{\rm sp}$, $\Omega_{\rm p}$, and $z_{\rm 0}$ 
are the amplitude of the spiral potential, the number of stellar spiral arms, 
the pitch angle, the pattern speed, and the scale height, respectively. 
In this study, we used $z_0 = 100$ pc and $R_0 = 1$ kpc. 
The spiral potential amplitude was controlled by the following dimensionless parameter: 
\begin{equation}
\mathcal{F} \equiv \frac{m|A|}{|\Phi_{\rm 0}| \sin i_{\rm sp}},
\end{equation}
which represents the gravitational force due to the spiral arms in the direction perpendicular to 
the arms relative to the radial force from the background axisymmetric potential \citep{Shu+1973}.
To model typical two-armed (i.e. $m = 2$) spiral galaxies, we used a typical $i_{\rm sp}$ of $25^\circ$,
which was determined based on observations of external spiral galaxies \citep[e.g.][]{Grosbol+2004}.
Since a gravitational potential is not a directly observed value, we used $\mathcal{F} = 5\%$ 
because of the results of theoretical studies of self-consistent steady spiral models \citep[e.g.][]{Grosbol1993}\footnote{
According to studies of self-consistent steady spiral models, the effects of nonlinear stellar dynamics cause
self-consistent spirals to terminate at the 4/1 resonance and would damp self-consistent spirals
for $\mathcal{F} \gtrsim 5\%$ \citep{ContopoulosGrosbol1988,Patsis+1991}.
}.
Although the value of $\Omega_{\rm p}$ is arbitrary and varies among galaxies, 
the aforementioned self-consistent steady spiral models suggest that a steady spiral can extend up to or beyond its CR radius. 
To analyse the gas velocity patterns at the same radii as those in DYNAMIC model (i.e. $R > 6$ kpc; see Section \ref{sec:method:dynamic}), 
we thus used $R_{\rm CR} \simeq 15$ kpc, which corresponds $\Omega_{\rm p} \simeq 16~\rm km~s^{-1}~kpc^{-1}$ 
in the STEADY model.

%%%%%%%%%%%%%%%%
\section{Gas flows and velocity patterns around spiral arms}
\label{sec:results}

We present spatial distributions of the gas density and velocity around spiral arms in the STEADY and DYNAMIC models.
In Section \ref{sec:results:distribution}, 
we examine the differences between the spatial distributions and gas flows in the two spiral models.
These results support those presented in \citet{Baba+2015a}.
Next we present the similarities and differences between the velocity patterns of the gas 
in the STEADY and DYNAMIC spiral models 
in Sections \ref{sec:results:velocitypattern:steady} and \ref{sec:results:velocitypattern:dynamic}, respectively.

%%%%%%%%%%%%%%%%
\subsection{Spatial distributions and flows of gas}
\label{sec:results:distribution}

\begin{figure*}
\begin{center}
\includegraphics[width=0.95\textwidth]{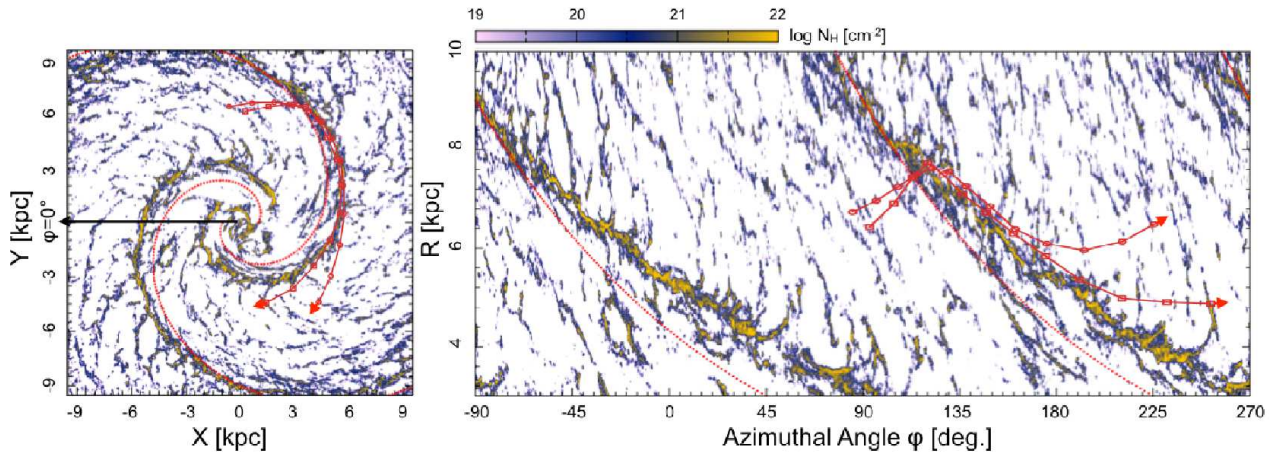}
\caption{
Spatial distributions of molecular gas in the $x-y$ plane (left) and the $R-\phi$ plane (right) in the STEADY model at $t=$ 360 Myr.
Solid lines with symbols indicate SPH particle trajectories. Dotted red lines indicate minima of of spiral potential.
Time evolutions are shown in a rotating frame of the spiral arm.
}	
\label{fig:DensityWaveSnapshot}
\end{center}
\end{figure*}

\begin{figure*}
\begin{center}
\includegraphics[width=0.95\textwidth]{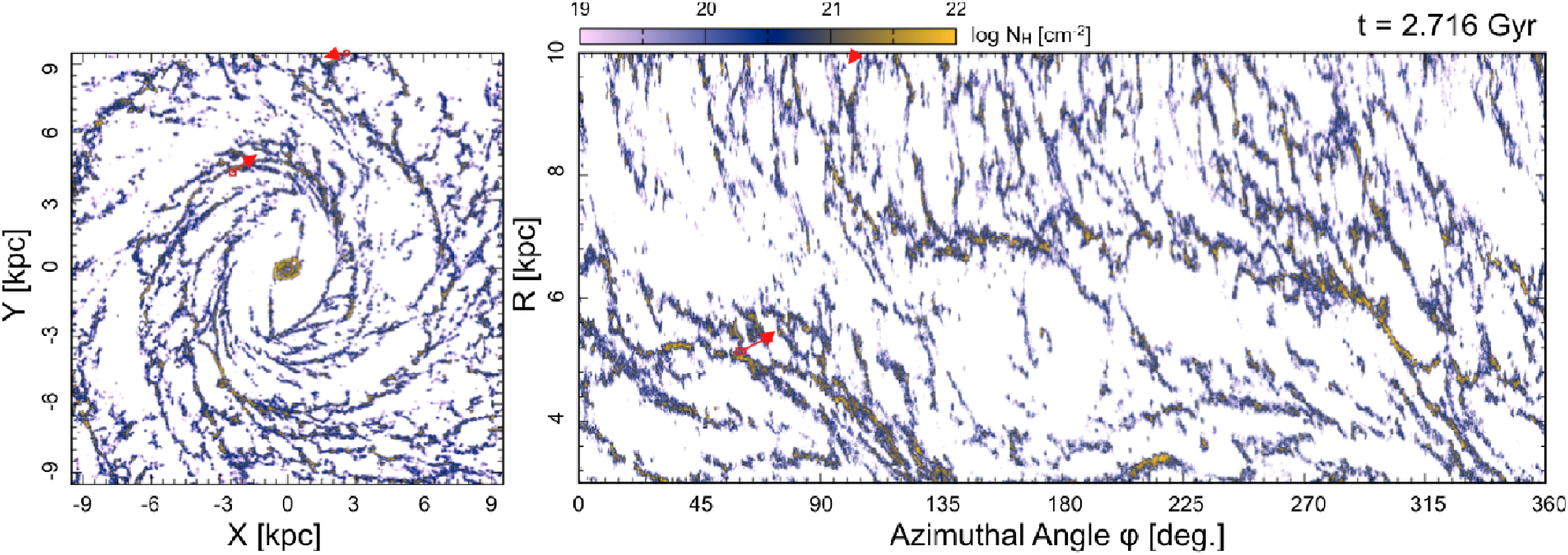}
\includegraphics[width=0.95\textwidth]{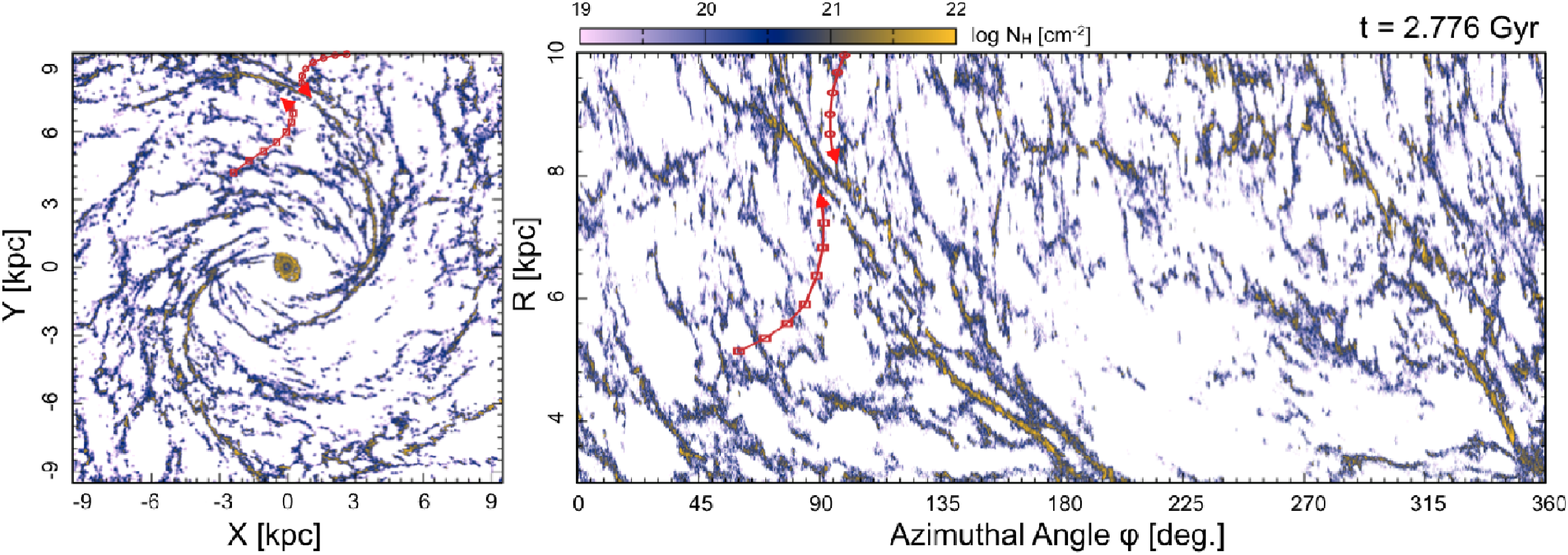}
\includegraphics[width=0.95\textwidth]{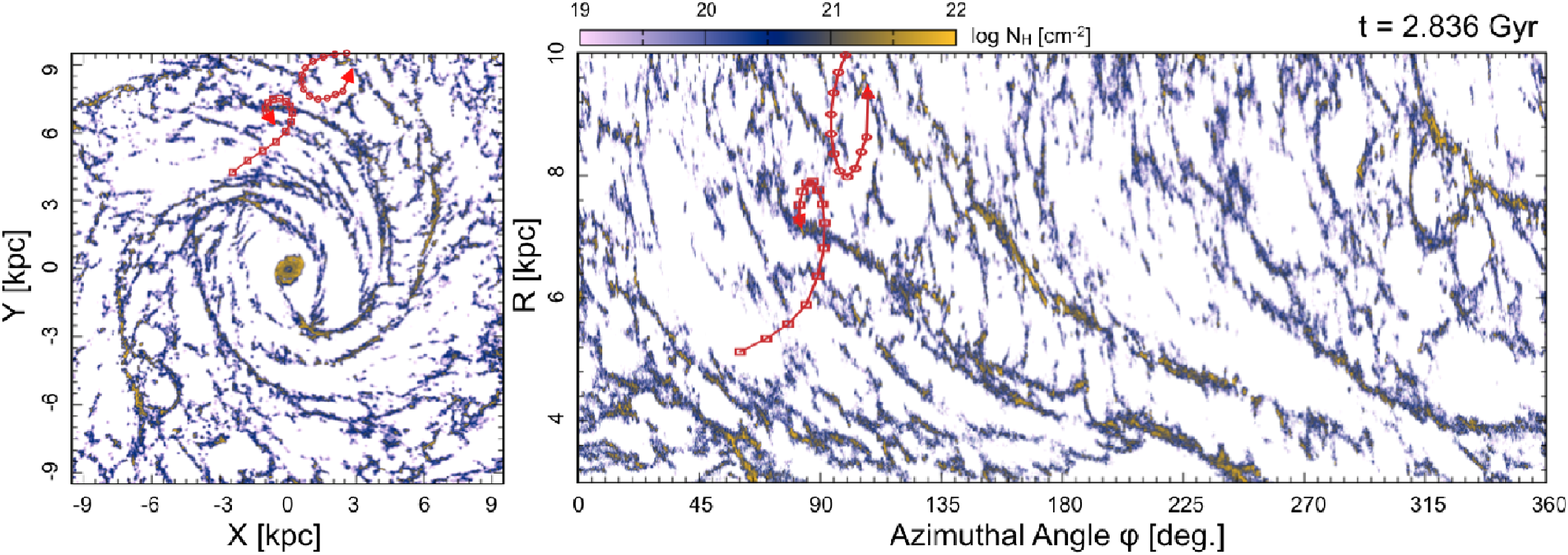}
\caption{
Same as Fig. \ref{fig:DensityWaveSnapshot}, but for DYNAMIC model at $t=$ 2.716 Gyr, 2.776 Gyr, and 2.836 Gyr, 
from top to bottom. Red lines with symbols in the $x-y$ plane indicate SPH particle trajectories.
Time evolution is shown in frame rotating at $\Omega = 25~\rm km~s^{-1}~kpc^{-1}$, 
which corresponds to galactic rotational angular frequency at $R \simeq 8$ kpc.
Note that in the inner part of the disc ($R \lesssim 3$ kpc)  the simulated galaxy shows also the characteristic features, 
such as offset ridges and inner rings, which are formed by the stella bars \citep[e.g.][]{Athanassoula1992b,Byrd+2006}.
}	
\label{fig:DynamicSpiralSnapshot}
\end{center}
\end{figure*}

Fig. \ref{fig:DensityWaveSnapshot} shows the surface density distributions of $\rm H_2$ gas in the STEADY model.
Molecular gas forms the main two-arm spirals along the (stellar) spiral arms.
There are also many substructures, i.e. `feathers' or `spurs', between the spirals and on the downstream sides of the spirals, 
as observed in actual spiral galaxies \citep[e.g.][]{Elmegreen1980,LaVigne+2006}. 
Because $R_{\rm CR} = 15$ kpc, the gas shown in these panels overtakes the spiral arms. 
For purpose of demonstration, the trajectories of two SPH particles are plotted (as red lines with symbols)
in the left panel; as shown, the SPH particles flow into the spiral from the trailing side, and experience 
a sudden change in direction due to galactic shocks, and then pass through the spiral onto the leading side.

\begin{figure}
\begin{center}
\includegraphics[width=0.48\textwidth]{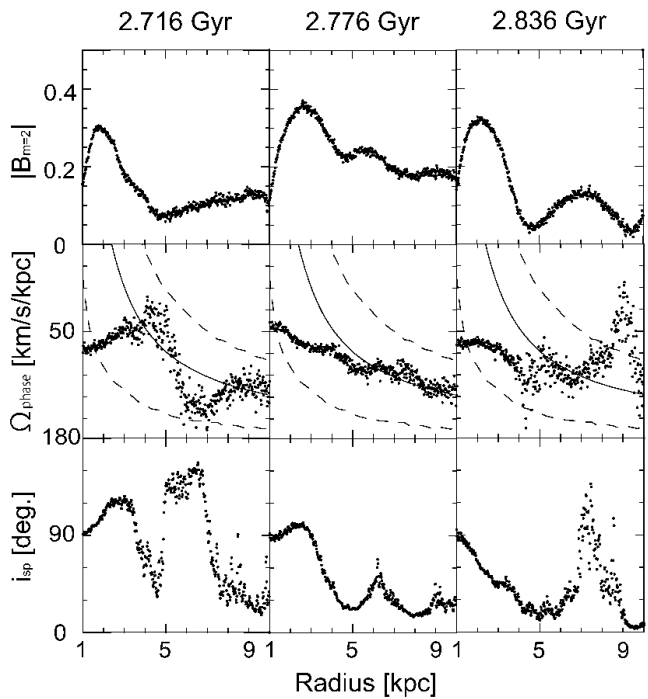}
\caption{
Radial profiles of the grand-design spiral arm parameters ($m=2$) in DYNAMIC model 
at times corresponding to snapshots shown in Fig. \ref{fig:DynamicSpiralSnapshot}.
Top: spiral amplitude $|B_{m=2}|$. 
Middle: angular phase speed $\Omega_{\rm phase}~\rm [km~s^{-1}~kpc^{-1}]$.
Solid and dashed curves indicate circular rotational frequency $\Omega_{\rm cir}$ and 
$\Omega_{\rm cir} \pm \kappa/2$ (here, $\kappa$ is epicyclic frequency), respectively.
Bottom: pitch angle $i_{\rm sp}$ [degrees].
These spiral parameters were analysed using 1D Fourier decomposition of 
stellar surface density distributions with respect to azimuthal direction at each radius
\citep[See Section 2.3 of][]{Baba2015c}.
}	
\label{fig:DynamicSpiralModeEvolution}
\end{center}
\end{figure}

The situation differs in the DYNAMIC model, for which the the surface density distributions of the molecular gas 
are shown in Fig. \ref{fig:DynamicSpiralSnapshot}.
The molecular spiral arms are clear at $t=2.776$ Gyr, but are weak at the other times. 
Such time evolution originates from the evolution of the stellar spirals.  
As shown Fig. \ref{fig:DynamicSpiralModeEvolution}, the amplitudes ($|B_{m=2}|$), 
rotational frequencies ($\Omega_{\rm phase}$), and pitch angles ($i_{\rm sp}$) of the spiral arms are not constant, 
but change within $\sim 100$ Myr (i.e. $\lesssim$ the typical rotational period of a galaxy).
In other words, the grand-design (stellar) spiral arms in the DYNAMIC model are not stationary, 
but rather transient recurrent \citep[i.e. dynamic;][]{Baba2015c}.
To compare the gas flow with that in the STEADY model, 
the SPH particle trajectories were overlaid on the gas density maps, 
as shown in Fig. \ref{fig:DynamicSpiralSnapshot}. 
In contrast to the STEADY model, in the DYNAMIC model, 
the SPH particle trajectories that are overlaid on the $x$-$y$ map of the gas show 
that gas and stars fall into the spiral arm from both sides, rather than from just one side
\citep[see also][]{DobbsBonnell2008,Wada+2011,Kawata+2014}\footnote{
In this paper, we focus on gas motions, but similar behaviour of stars has been observed for stars in previous simulation results 
\citep{SellwoodBinney2002,Grand+2012a,Grand+2012b,Roskar+2012,Baba+2013,Grand+2014,Kawata+2014}.}.

\begin{figure*}
\begin{center}
\includegraphics[width=0.45\textwidth]{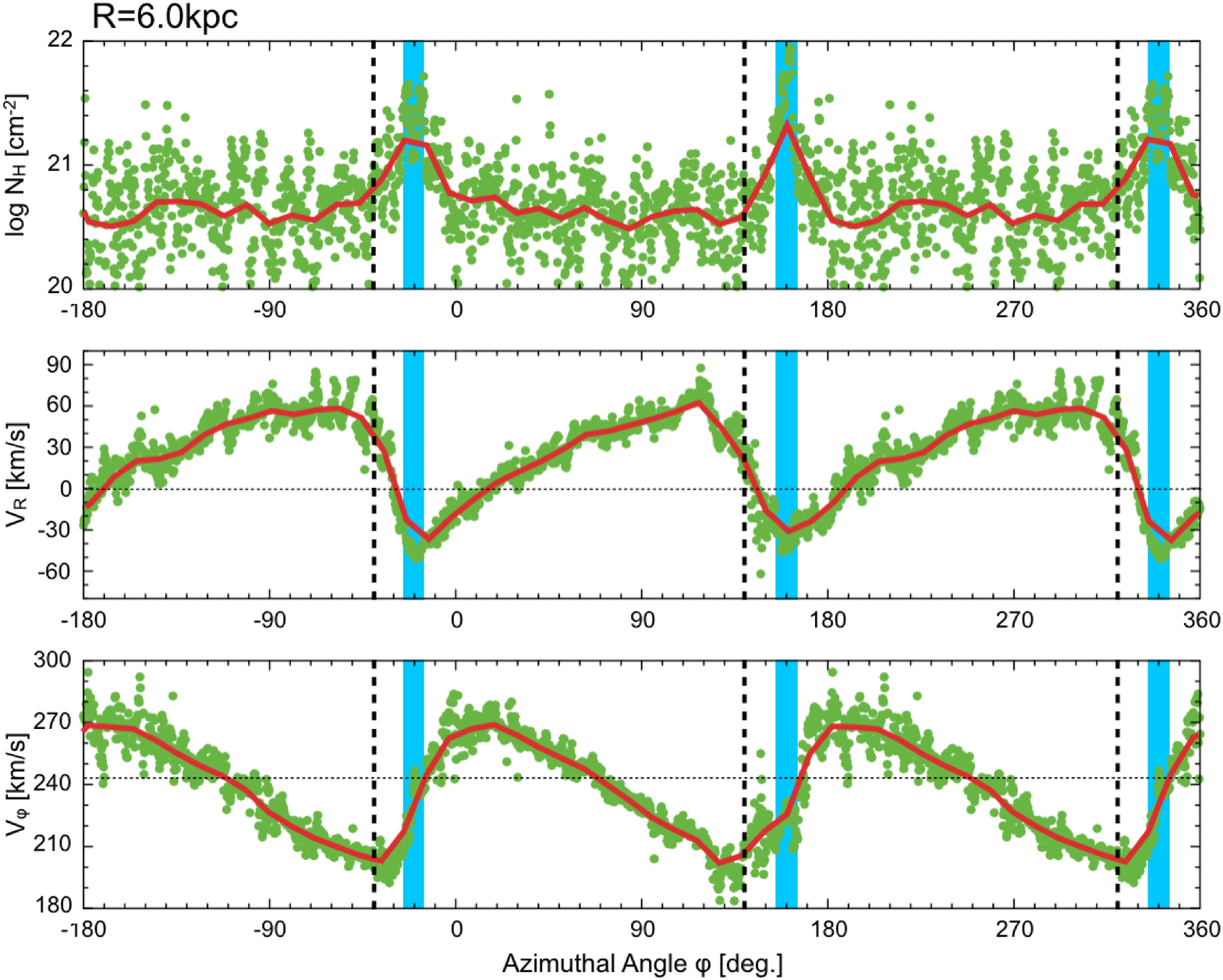}
\includegraphics[width=0.45\textwidth]{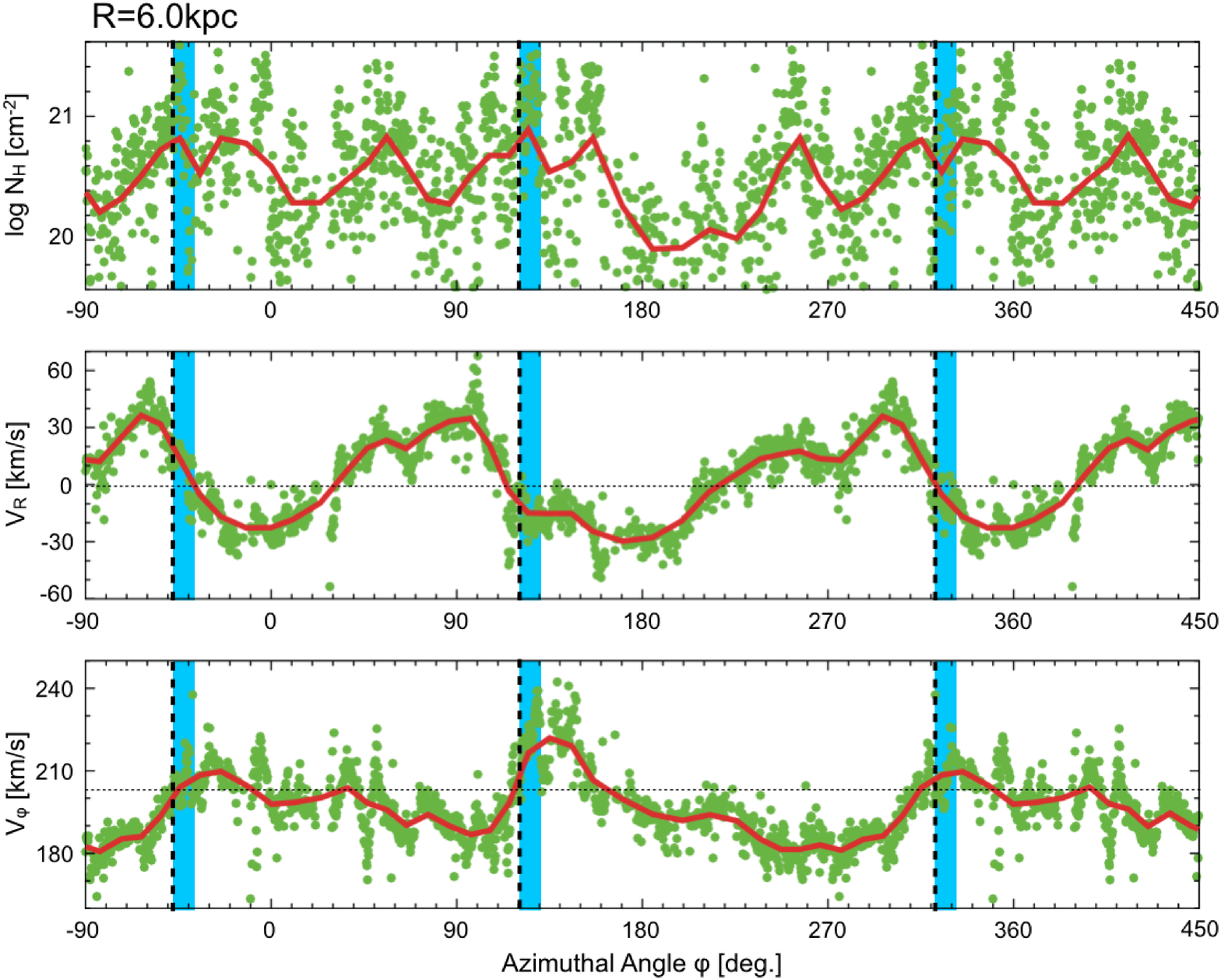}

\includegraphics[width=0.45\textwidth]{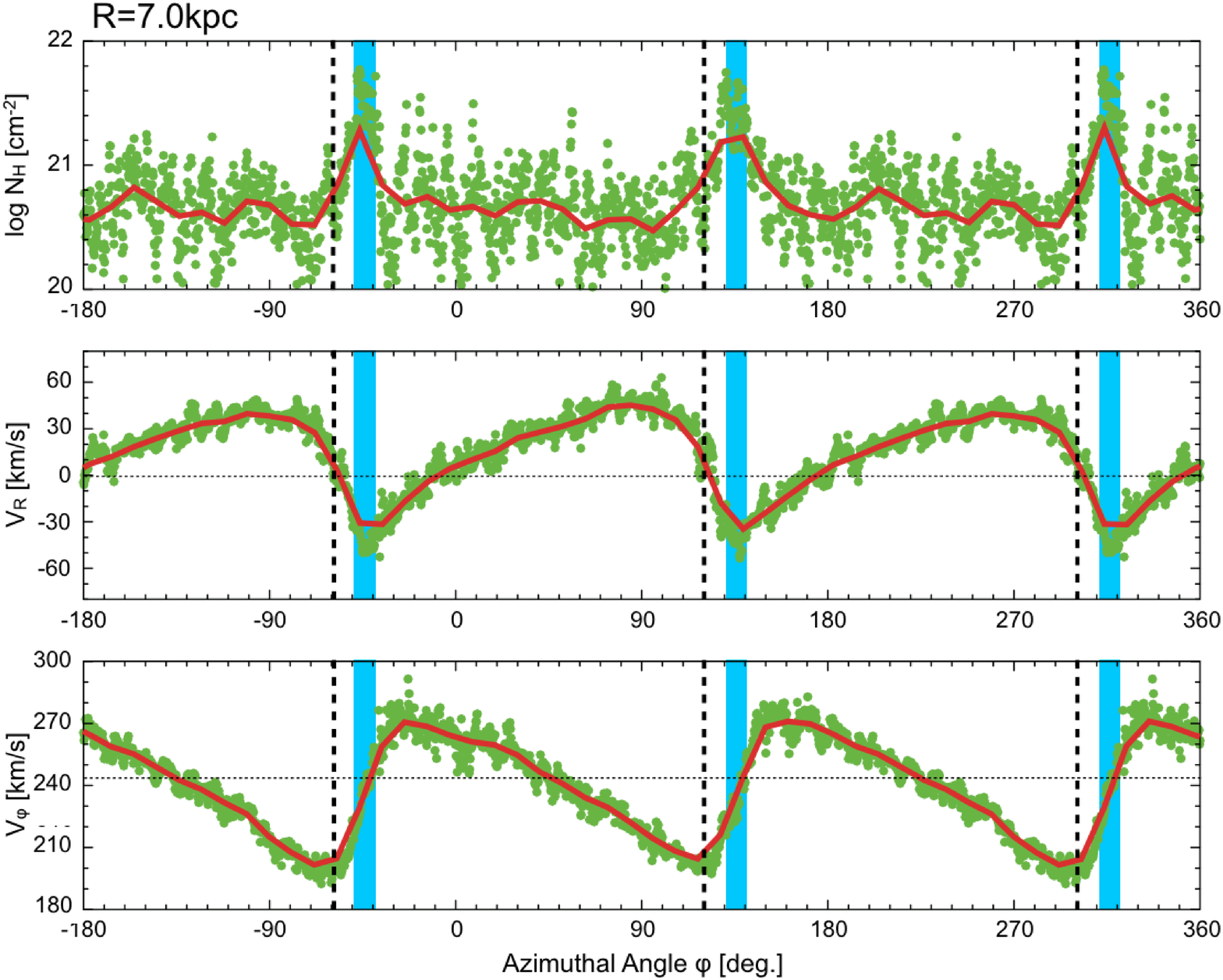}
\includegraphics[width=0.45\textwidth]{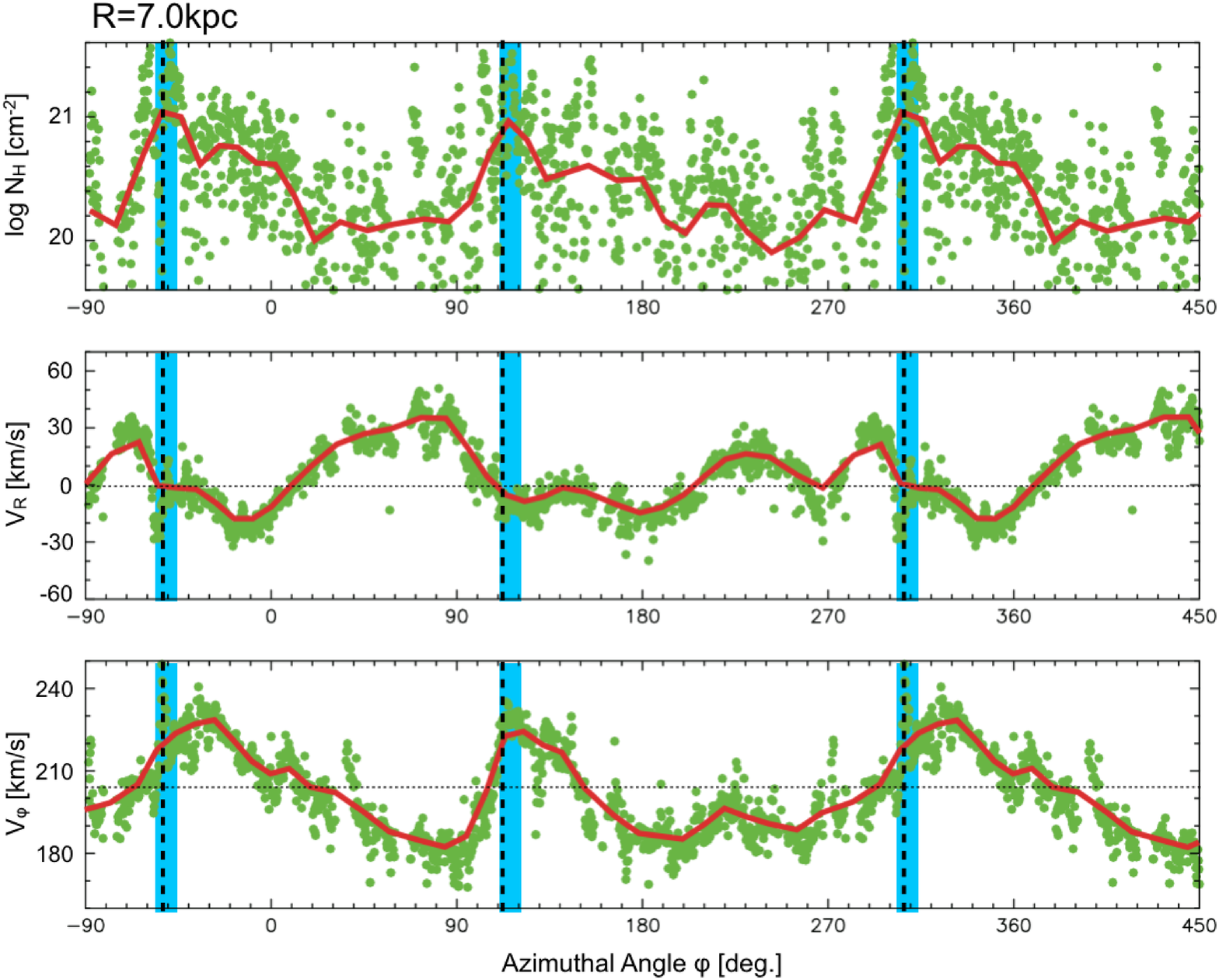}

\includegraphics[width=0.45\textwidth]{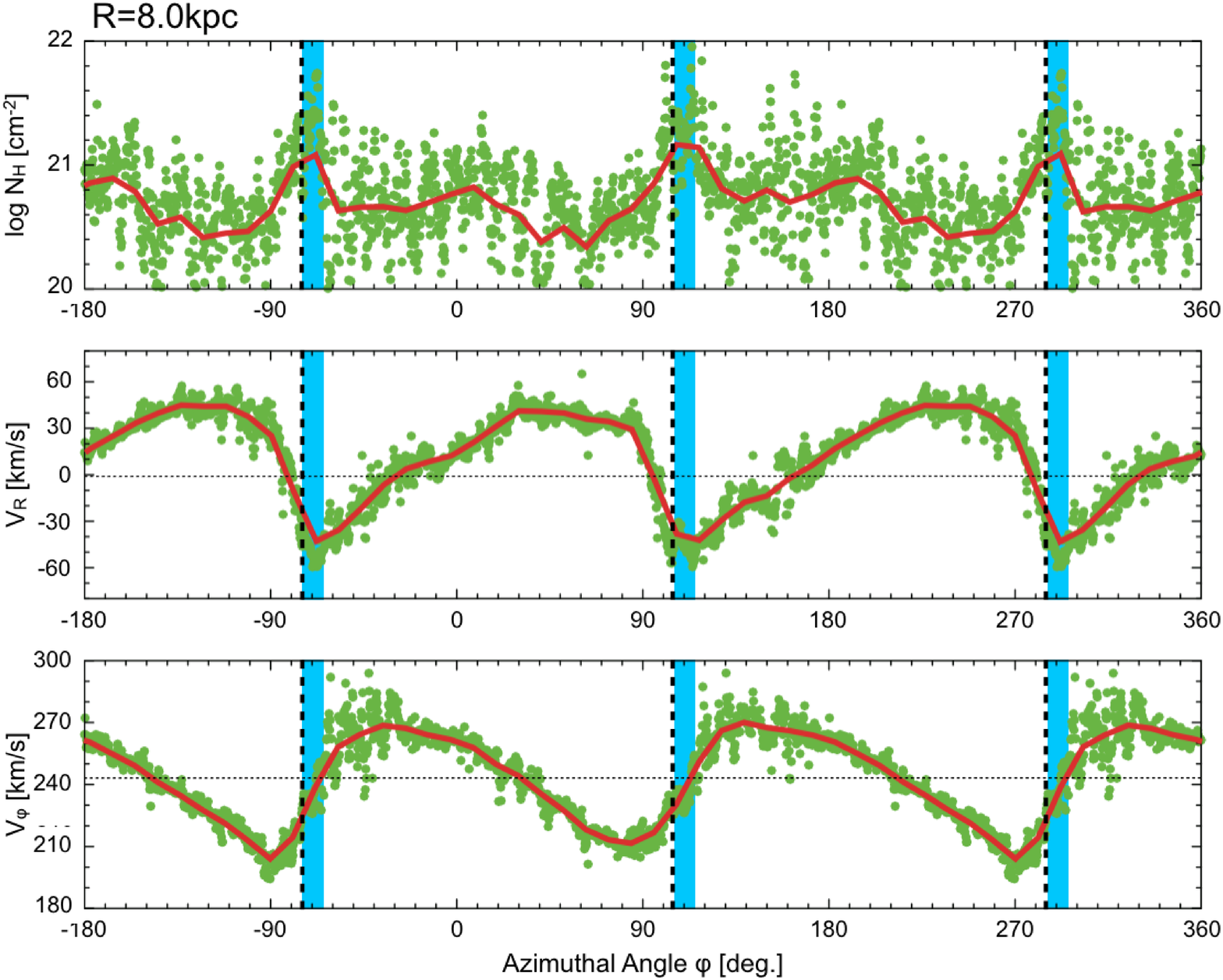}
\includegraphics[width=0.45\textwidth]{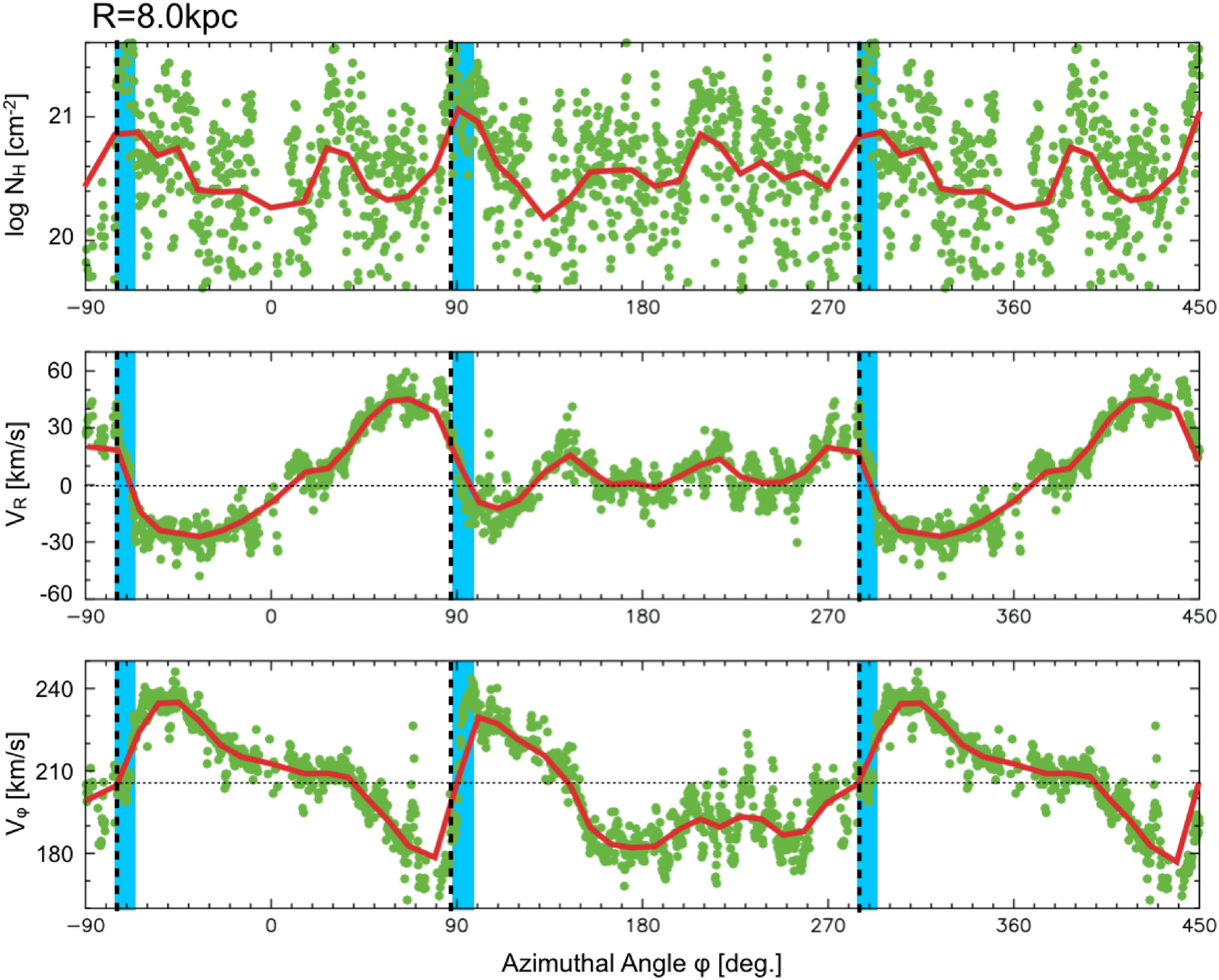}
\caption{
Left: Azimuthal profiles of $N_{\rm H}$ (top), $V_R$ (middle) and $V_\phi$ (bottom) of gas 
in STEADY model (shown in Fig. \ref{fig:DensityWaveSnapshot}) at $R = 6.0$ kpc, $7.0$ kpc, and $8.0$ kpc. 
Gas flows from left ($\phi<0$) to right ($\phi>0$) for the STEADY model.
Solid curves fit averaged values over 5 Myr using a cubic spline interpolation.
Horizontal dotted lines indicate average velocities at each radius, respectively.
Vertical black dashed lines and vertical blue shaded regions indicate 
positions of spiral potential minimum and gas density maximum, respectively.
Right: Same as left panels, but for DYNAMIC model 
(shown in Fig. \ref{fig:DynamicSpiralSnapshot}) at $R= 6.0$ kpc, $7.0$ kpc, and $8.0$ kpc.
Time is $t = 2.776$ Gyr.
}	
\label{fig:AzimuthProfiles}
\end{center}
\end{figure*}

The left panels of Fig. \ref{fig:AzimuthProfiles} show that, in the STEADY model, 
the primary gas density peaks (i.e. gaseous spiral arms) occur on the downstream sides of the spiral potential minima 
at all presented radii (see also the right panel of Fig. \ref{fig:DensityWaveSnapshot})\footnote{
The secondary gas density peaks also occur downstream from the primary peaks,  
e.g. $\phi \simeq 260^\circ$, $200^\circ$, and $190^\circ$ at $R = 6.0$, 7.0, and 8.0 kpc, respectively, 
and correspond to the spurs (see the right panel of Fig. \ref{fig:DensityWaveSnapshot}).
The existence of such secondary peaks was previously observed 
in early galactic shock calculations \citep[e.g.][]{Roberts1969,Shu+1973}, 
although the origins of the secondary peaks are beyond the scope of this paper. 
See \citet{DobbsBaba2014} and the references therein for discussions of their origins.
}.
Such radial dependence of the positions of the gaseous arms relative to spiral arms has been indicated 
by previous hydrodynamic simulations of steady spiral models \citep[e.g.][]{GittinsClarke2004,Baba+2015a}.
In contrast, the DYNAMIC model (right panels of Fig. \ref{fig:AzimuthProfiles}) shows no systematic offset 
between the gaseous and spiral arms. The same results have been reported in previous studies \citep[e.g.][]{Baba+2015a}.

%%%%%%%%%%%%%%%
\subsection{Streaming motions in steady spiral model}
\label{sec:results:velocitypattern:steady}

The azimuthal profiles of the gas velocities in the STEADY model are shown in the left panels of Fig. \ref{fig:AzimuthProfiles}.
The vertical blue shaded regions indicate the positions of the gaseous spiral arms. 
Focusing on the streaming motions of the gaseous spiral arms, it is evident that 
the gaseous spiral arms are associated with the $V_R$ minima and $V_\phi \simeq \bar{V}_\phi$, 
where $\bar{V}_\phi$ is the average rotational velocity, i.e. the galactic rotational velocity.
Such streaming patterns do not depend on the pitch angle of the steady spiral within the range of 
observed pitch angles $10^\circ \lesssim i_{\rm sp} \lesssim 40^\circ$ \citep[e.g.][]{Grosbol+2004}.
Therefore, it is suggested that if an observed spiral galaxy has steady stellar spiral arms, 
the gas in the {\it gaseous} spiral arms is expected to exhibit strong {\it radial} streaming motions
and weak tangential streaming motions, at least well inside $R_{\rm CR}$.

This result is consistent with the predictions of the linear density wave theory \citep{Lin+1969,Burton1971} and
the QSSS/galactic shock hypothesis \citep[i.e. non-linear density wave theory; cf. Fig. 5 of][]{Roberts1969}.
More specifically, the QSSS/galactic shock hypothesis predicts that $V_R$ steeply decreases toward 
the location of maximum gas compression (i.e. the gaseous spiral arm), 
reaches its minimum, and then increases on the outer side of the arm. 
Furthermore, this theory predicts that $V_\phi$ decrease just before the gas enters the stellar spiral arm, 
and then increases steeply in the arm. 
Therefore, gaseous arms are located to be at a transition points from a $V_\phi$ minima to a $V_\phi$ maxima.

In contrast to the patterns in gaseous arms, the gas streaming in the {\it stellar} spirals are radius-dependent. 
From the left panels of Fig. \ref{fig:AzimuthProfiles}, 
it is evident that when $R = 6.0$ kpc, the gases in the stellar spirals 
(represented by vertical dashed lines) have $V_R \simeq 0$ as well as $V_\phi$ minima, 
whereas when $R=8.0$ kpc they have the $V_R$-minima and $V_\phi \simeq \bar{V}_\phi$.
These differences result from the radial dependences of the positions of 
gaseous arms (vertical blue shaded regions) relative to stellar spirals (vertical dashed lines).

%%%%%%%%%%%%%%%
\subsection{Streaming motions in dynamic spiral model}
\label{sec:results:velocitypattern:dynamic}

The right panels of Fig. \ref{fig:AzimuthProfiles} show the azimuthal gas velocity profiles in the DYNAMIC model.
Although the gas flows with respect to the spiral arms in the DYNAMIC model are 
completely different from those in the STEADY model (see Section \ref{sec:results:distribution}), 
the velocity profiles of the two models are similar.
This similarity is due to the convergence of gas from both side of a spiral arm.
The gas whose guiding centre is at an inner (outer) radius enters a spiral arm from behind (in front of) the arm, and then 
form a clear grand-design spiral at $t \simeq 2.776$ Gyr (right panels of Fig. \ref{fig:DynamicSpiralModeEvolution}).
These directions of the motions are due to epicyclic motion\footnote{
In general, the phrase `epicyclic motion' refers to near-circular motion \citep[][]{BinneyTremaine2008}. 
However, in this paper, we also use this phrase to describe non-circular orbits with finite radial amplitudes.}, 
whose directions are opposite to the galactic rotation direction.
In other words, in order to conserve angular momentum, $V_\phi$ must be larger 
when the star or gas is closer to the galactic centre than when it is farther away.
In this case, the gas behind (in front of) a spiral arm tends to have velocities of 
$V_R <0$ ($V_R >0$) and $V_\phi \gtrsim {\bar V}_{\phi}$ ($V_\phi \lesssim {\bar V}_{\phi}$).

Although the overall velocity patterns in the DYNAMIC model are similar to those of the STEADY model, the details differ.
Focusing on the streaming motions of the {\it gaseous} arms in the DYNAMIC model, it is evident that 
the $V_R$ minima are not associated with the gaseous arms; instead, the gaseous arms have velocities of $V_R \simeq 0$ 
and ${\bar V}_{\phi} \lesssim V_\phi \lesssim V_{\phi,\rm max}$ (right panels of Fig. \ref{fig:AzimuthProfiles}).
These velocity patterns result from the fact that stellar spiral arms are formed by flows from both sides of the arms, 
and that gas is associated with this accumulation process (see below).

\begin{figure}
\begin{center}
\includegraphics[width=0.48\textwidth]{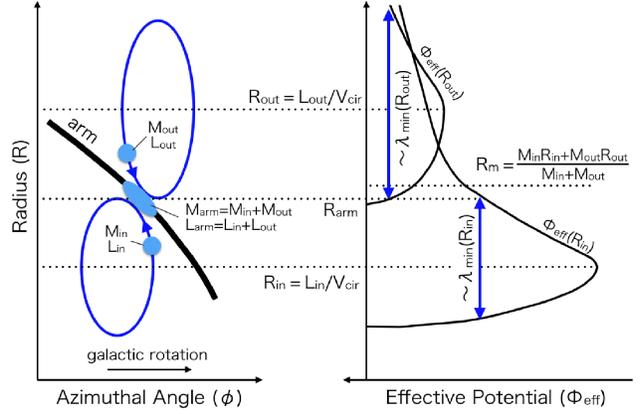}
\caption{
Schematic of dynamic spiral model, where 
spiral arm is formed by collision between epicyclic flows with guiding centres at $R_{\rm in}$ and $R_{\rm out}$.
Details are described in Section \ref{sec:results:velocitypattern:dynamic}.
}	
\label{fig:DynamicSpiralModeSchematic}
\end{center}
\end{figure}

In order to explain why these velocity patterns occur in the dynamic spiral model, 
we consider a simple model in which a galaxy has a flat rotation curve at velocity $V_{\rm cir}$ and 
a spiral arm is formed by the collision between epicyclic flows with guiding centres 
at an inner radius ($R_{\rm in}$) and an outer radius ($R_{\rm out}$).
The masses of the flows from the inner and outer radii are assumed to be $M_{\rm in}$ and $M_{\rm out}$, respectively.
A schematic of this simple model is presented in Fig. \ref{fig:DynamicSpiralModeSchematic}.
Assuming angular momentum conservation, the angular momentum of the arm is given by 
$L_{\rm arm} = M_{\rm in} R_{\rm in} V_{\rm cir} + M_{\rm out} R_{\rm out} V_{\rm cir}$,
and the mass of the arm is $M_{\rm arm} = M_{\rm in} + M_{\rm out}$.
Thus, the rotational velocity of the arm ($V_{\rm arm}$) formed at $R_{\rm arm}$ is given by
\begin{eqnarray}
 V_{\rm arm} = \frac{L_{\rm arm}}{M_{\rm arm} R_{\rm arm}} 
 = \frac{M_{\rm in} R_{\rm in} + M_{\rm out} R_{\rm out}}{(M_{\rm in} + M_{\rm out})R_{\rm arm}} V_{\rm cir}
 = \frac{R_{\rm m}}{R_{\rm arm}} V_{\rm cir},
\end{eqnarray}
where $R_{\rm m} \equiv (M_{\rm in} R_{\rm in} + M_{\rm out} R_{\rm out})/(M_{\rm in} + M_{\rm out})$ 
is the mass-weighted average radius of the raw material forming the arm, and can be rewritten as follows
\begin{eqnarray}
 R_{\rm m} 
 = \frac{M_{\rm out}(R_{\rm out}-R_{\rm arm}) - M_{\rm in}(R_{\rm arm}-R_{\rm in})}{M_{\rm in} + M_{\rm out}} + R_{\rm arm}.
\end{eqnarray}

To calculate $R_{\rm m}$, we first consider the simple case of $M_{\rm in} = M_{\rm out}$. 
In this case, 
\begin{eqnarray}
 R_{\rm m} = \frac{(R_{\rm out}-R_{\rm arm}) - (R_{\rm arm}-R_{\rm in})}{2} + R_{\rm arm}.
\end{eqnarray}
Typically, the epicyclic amplitude is approximated by the wavelength at the minimum frequency 
of the Lin--Shu--Kalnajs dispersion relation $\lambda_{\rm min} = \lambda_{\rm crit} Q$,
where $\lambda_{\rm crit}$ is Toomre's critical wavelength and $Q$ is the Toomre's $Q$ parameter\footnote{
Considering a disc with the radial velocity dispersion ($\sigma_R$) and epicyclic frequency ($\kappa$), 
a typical epicyclic amplitude $\Delta R$ is approximated by $\sigma_R/\kappa$ \citep[e.g.][]{Bertin2000}.
On the other hand, $\lambda_{\rm min} = \lambda_{\rm crit} Q = 4 \sigma_R/\kappa$
\citep[see equation (8) and Fig. 2 of][]{DobbsBaba2014}. Thus, $\Delta R \sim \lambda_{\rm min}$.
}.
Thus,
\begin{eqnarray}
 R_{\rm out} - R_{\rm arm} \approx \lambda_{\rm min} (R_{\rm out})
\end{eqnarray}
and
\begin{eqnarray}
 R_{\rm arm} - R_{\rm in} \approx \lambda_{\rm min} (R_{\rm in}).
\end{eqnarray}
Since dynamic spiral arms develop from a structure with a wavelength of $\lambda_{\rm crit}$
via the swing amplification mechanism \citep[e.g.][]{CarlbergFreedman1985,Fujii+2011,Baba+2013},
$\lambda_{\rm crit}$ can be approximated as follows
\begin{eqnarray}
 \lambda_{\rm crit}(R)  = \frac{2 \pi R}{m X_{\rm GT}} \approx 2 R,
\end{eqnarray}
where $m$ is the number of arms, 
and $X_{\rm GT}$ is Goldreich \& Tremaine's parameter \citep[see equation (19) of][]{DobbsBaba2014}; 
$m = 2$ and $X_{\rm GT} \approx 1.5$ for maximum amplification 
\citep[see Fig. 5 of][and see also \citet{MichikoshiKokubo2016b}]{DobbsBaba2014}.
Thus, if $Q(R_{\rm out}) \approx Q(R_{\rm in}) \approx 1$, 
\begin{eqnarray}
 R_{\rm m} 
  \approx \frac{\lambda_{\rm min}(R_{\rm out}) - \lambda_{\rm min}(R_{\rm in})}{2} + R_{\rm arm} \nonumber\\
  \approx (R_{\rm out} - R_{\rm in}) + R_{\rm arm} > R_{\rm arm}.
\end{eqnarray}
That is $R_{\rm m}/R_{\rm arm} > 1$, resulting in $V_{\rm arm} > V_{\rm cir}$ from equation (3).

However, $M_{\rm in}$ might be larger than $M_{\rm out}$ 
because the radial mass distributions in disc galaxies tend to follow exponential profiles.
In such cases, the difference between $M_{\rm in} (R_{\rm arm} - R_{\rm in})$ and $M_{\rm out} (R_{\rm out} - R_{\rm arm})$
tends to decrease, resulting in the condition $V_{\rm arm} \simeq V_{\rm cir}$.
It is therefore suggested that if an observed spiral galaxy has a dynamic spiral, 
then the {\it gaseous} arms should exhibit streaming motions with $V_{\rm arm} \gtrsim V_{\rm cir} \gtrsim \bar{V}_\phi$.

Finally, we discuss the gas streaming motions in the stellar spirals. 
Unlike in the STEADY model, in the DYNAMIC model, the gas streaming motions 
are nearly zero in the {\it stellar} arms regardless of the radius,
because of the small offsets between gaseous arms and stellar arms.

%%%%%%%%%%%%%%%%
\section{Summary and Implications}
\label{sec:discussion}

In this paper, we analysed the results of hydrodynamic simulations 
and discussed  the differences between the velocity patterns predicted by the steady and dynamic spiral models.
The main results can be summarized as follows. The steady spiral model shows that 
\begin{enumerate}
\item the locations of gaseous spiral arms move monotonically from downstream to 
upstream of the stellar spiral arms with increasing radius \citep[see also][]{Baba+2015a}; 
\item gaseous arms well inside the $R_{\rm CR}$ are associated with $V_R$ minima and 
weak tangential streaming motions, i.e. {\it radial} streaming motions; 
\item gas streaming patterns in the stellar spiral arms are radius-dependent,
because of the radial dependence of the gaseous arms position relative to the stellar arm positions.
\end{enumerate}
These results are consistent with the predictions of conventional galactic shock theory \citep{Roberts1969,Lin+1969,Burton1971}.
However, at least in terms of our simulations, the dynamic spiral simulation results suggest that 
\begin{enumerate}
\item no systematic offset exists between the gaseous arm and the stellar arms \citep[see also][]{Baba+2015a}; 
\item gaseous arms tend to associate with $V_R \simeq 0$ and $V_\phi \gtrsim \bar{V}_\phi$,
i.e. {\it tangential} streaming motions; 
\item gas streaming motions are nearly zero in the stellar spiral arms.
\end{enumerate}

These results were obtained in the hydrodynamic simulations of isolated galaxy models, 
although we expect the velocity patterns of tidally induced spirals (e.g. M51 and M81) 
to differ between the steady and dynamic spiral models. 
According to $N$-body/hydrodynamic simulations \citep{Dobbs+2010} and $N$-body simulations \citep{Oh+2008,Oh+2015} 
of tidally interacting systems, tidally-induced spirals are not quasi-stationary density waves, 
but {\it kinematic} density waves.
Furthermore, these simulations suggested that the pattern speeds of tidally induced spirals clearly 
differ from the galactic angular speed, but decrease as the radius increases \citep{Pettitt+2016}. 
This scenario illustrates intermediate behavior between the steady and dynamic spiral models.
In this case, the gas enters into a spiral arm from one side of the arm, 
and then experiences a sudden compression such as a galactic shock. 
Thus, it is expected that the gaseous spiral arms in tidally induced spiral galaxies also have radial streaming motions,
which is is consistent with previous observations of M51 \citep[e.g.][]{KunoNakai1997,Shetty+2007,Miyamoto+2014}.
However, in contrast to the steady spiral model (see Section \ref{sec:results:velocitypattern:steady}), 
the gas streaming velocities in {\it stellar} arms might not vary with radius,
because the positions of the gaseous arms relative to the tidally induced stellar spirals 
are not expected to be strongly radius-dependent.

These differences encourage the use of  gas velocity patterns in spiral galaxies for observational spiral model tests.
Our simulations showed that the streaming velocity is typically $\sim 10~\rm km~s^{-1}$
and that the offset between stellar and gaseous arms is typically $\lesssim 1$ kpc, suggesting that 
the required spatial and velocity resolutions are at least $\lesssim 1$ kpc and $\sim 10~\rm km~s^{-1}$, respectively.
In addition, this `velocity pattern method' requires gas detection in both the arm and inter-arm regions.
These spatial and velocity resolutions are easily achievable with existing instruments, 
although the gas detection sensitivity in inter-arm regions is somewhat problematic (except for nearest galaxies).
It is therefore useful to conduct high-sensitivity observations of the arm and inter-arm region 
with the latest generation of instruments such as the Atacama Large Millimeter/submillimeter Array (ALMA) 
and the Square Kilometre Array (SKA).

Nevertheless, an accurate method of determining gas velocity patterns in galaxies requires further study,
because it is only possible to measure the line-of-sight velocity of a gas. 
In other words, one cannot directly measure both $V_R$ and $V_\phi$ at the same point in a galaxy.
In fact, the conventional streaming velocity measurement method is based on analysing the position-velocity (PV) diagrams 
along the major and minor axes of the observed spiral galaxies \citep[e.g.][]{Aalto+1999}; 
tangential streaming should be most apparent along the major axis, 
whereas radial streaming should be most apparent along the minor axis. 
However, the PV diagram method can be used to determine {\it local} streaming motions only at the points 
at which the spiral arm lies across the major or minor axis. In short, the PV diagram method cannot 
be used to ascertain the {\it global} distributions of streaming motions in observed spiral galaxies. 
Thus, in order to determine these global distributions of streaming motions, 
the 2D gas velocity fields in spiral galaxies must be acquired using another method, 
such as that proposed proposed by \citet{KunoNakai1997} \citep[see also Fig. 10 of][]{Miyamoto+2014}.
Methods of modelling spiral galaxy 2D velocity fields based on observational data will be useful to discriminate 
between the spiral models and will be presented in future reports.

Finally, we note that the velocity patterns themselves are similar between the two spiral models
and suggest that such similarity indicates that the kinematic method should not be used 
to determine $R_{\rm CR}$ from residual velocity fields or streaming motion direction changes. 
In fact, if one applies this so-called `geometric phase method' \citep[][]{Canzian1993} to determine $R_{\rm CR}$ (or a pattern speed) 
for a spiral galaxy with {\it dynamic} spirals, an $R_{\rm CR}$ value greater than the disc radius will be obtained.
Furthermore, the similarity between the two spiral models suggests that the existence of streaming motions 
is not conclusive evidence of the steady spiral model (i.e. the QSSS/galactic shock hypothesis). 
Therefore, it is important to analyse the global gas velocity patterns,
along with CO-H$\alpha$ offset measurements \citep{Egusa+2009} and gas-star offset measurements \citep{Baba+2015a}
to distinguish the spiral models.

\section*{Acknowledgements}

We would like to thank the referees for their critical comments and
informative reports which improved this paper. 
We thank Takayuki R. Saitoh, Michiko S. Fujii and Keiichi Wada
for a careful reading of the manuscript and constructive comments.
Calculations, numerical analyses and visualization were carried out on Cray XC30, and computers 
at Center for Computational Astrophysics, National Astronomical Observatory of Japan.  
This research was supported by HPCI Strategic Program Field 5 `The origin of matter and the universe'
and JSPS Grant-in-Aid for Young Scientists (B) Grant Number 26800099.

%\bibliographystyle{apj}
%\bibliography{ms}

\begin{thebibliography}{120}
\expandafter\ifx\csname natexlab\endcsname\relax\def\natexlab#1{#1}\fi

\bibitem[{{Aalto} {et~al.}(1999){Aalto}, {H{\"u}ttemeister}, {Scoville}, \&
  {Thaddeus}}]{Aalto+1999}
{Aalto}, S., {H{\"u}ttemeister}, S., {Scoville}, N.~Z., \& {Thaddeus}, P. 1999,
  \apj, 522, 165

\bibitem[{{Athanassoula}(1992)}]{Athanassoula1992b}
{Athanassoula}, E. 1992, MNRAS, 259, 345

\bibitem[{{Baba}(2015)}]{Baba2015c}
{Baba}, J. 2015, \mnras, 454, 2954

\bibitem[{{Baba} {et~al.}(2009){Baba}, {Asaki}, {Makino}, {Miyoshi}, {Saitoh},
  \& {Wada}}]{Baba+2009}
{Baba}, J., {Asaki}, Y., {Makino}, J., {Miyoshi}, M., {Saitoh}, T.~R., \&
  {Wada}, K. 2009, \apj, 706, 471

\bibitem[{{Baba} {et~al.}(2015){Baba}, {Morokuma-Matsui}, \&
  {Egusa}}]{Baba+2015a}
{Baba}, J., {Morokuma-Matsui}, K., \& {Egusa}, F. 2015, \pasj, 67, L4

\bibitem[{{Baba} {et~al.}(2016){Baba}, {Morokuma-Matsui}, \&
  {Saitoh}}]{Baba+2016b}
{Baba}, J., {Morokuma-Matsui}, K., \& {Saitoh}, T.~R. 2016, submitted

\bibitem[{{Baba} {et~al.}(2013){Baba}, {Saitoh}, \& {Wada}}]{Baba+2013}
{Baba}, J., {Saitoh}, T.~R., \& {Wada}, K. 2013, \apj, 763, 46

\bibitem[{{Bertin}(2000)}]{Bertin2000}
{Bertin}, G. 2000, {Dynamics of Galaxies}

\bibitem[{{Bertin} \& {Lin}(1996)}]{BertinLin1996}
{Bertin}, G., \& {Lin}, C.~C. 1996, {Spiral structure in galaxies a density
  wave theory}, ed. G.~{Bertin} \& C.~C. {Lin}

\bibitem[{{Bertin} {et~al.}(1989{\natexlab{a}}){Bertin}, {Lin}, {Lowe}, \&
  {Thurstans}}]{Bertin+1989a}
{Bertin}, G., {Lin}, C.~C., {Lowe}, S.~A., \& {Thurstans}, R.~P.
  1989{\natexlab{a}}, \apj, 338, 78

\bibitem[{{Bertin} {et~al.}(1989{\natexlab{b}}){Bertin}, {Lin}, {Lowe}, \&
  {Thurstans}}]{Bertin+1989b}
---. 1989{\natexlab{b}}, \apj, 338, 104

\bibitem[{{Binney} \& {Tremaine}(2008)}]{BinneyTremaine2008}
{Binney}, J., \& {Tremaine}, S. 2008, {Galactic Dynamics: Second Edition}
  (Princeton University Press)

\bibitem[{{Block} {et~al.}(1994){Block}, {Bertin}, {Stockton}, {Grosbol},
  {Moorwood}, \& {Peletier}}]{Block+1994}
{Block}, D.~L., {Bertin}, G., {Stockton}, A., {Grosbol}, P., {Moorwood},
  A.~F.~M., \& {Peletier}, R.~F. 1994, \aap, 288, 365

\bibitem[{{Burton}(1971)}]{Burton1971}
{Burton}, W.~B. 1971, \aap, 10, 76

\bibitem[{{Byrd} {et~al.}(2006){Byrd}, {Freeman}, \& {Buta}}]{Byrd+2006}
{Byrd}, G.~G., {Freeman}, T., \& {Buta}, R.~J. 2006, \aj, 131, 1377

\bibitem[{{Byrd} \& {Howard}(1992)}]{ByrdHoward1992}
{Byrd}, G.~G., \& {Howard}, S. 1992, \aj, 103, 1089

\bibitem[{{Canzian}(1993)}]{Canzian1993}
{Canzian}, B. 1993, \apj, 414, 487

\bibitem[{{Carlberg} \& {Freedman}(1985)}]{CarlbergFreedman1985}
{Carlberg}, R.~G., \& {Freedman}, W.~L. 1985, \apj, 298, 486

\bibitem[{{Cepa} \& {Beckman}(1990)}]{CepaBeckman1990}
{Cepa}, J., \& {Beckman}, J.~E. 1990, \apj, 349, 497

\bibitem[{{Contopoulos} \& {Grosbol}(1988)}]{ContopoulosGrosbol1988}
{Contopoulos}, G., \& {Grosbol}, P. 1988, \aap, 197, 83

\bibitem[{{Dobbs} \& {Baba}(2014)}]{DobbsBaba2014}
{Dobbs}, C., \& {Baba}, J. 2014, PASA, 31, 35

\bibitem[{{Dobbs} \& {Bonnell}(2008)}]{DobbsBonnell2008}
{Dobbs}, C.~L., \& {Bonnell}, I.~A. 2008, \mnras, 385, 1893

\bibitem[{{Dobbs} \& {Pringle}(2010)}]{DobbsPringle2010}
{Dobbs}, C.~L., \& {Pringle}, J.~E. 2010, \mnras, 409, 396

\bibitem[{{Dobbs} {et~al.}(2014){Dobbs}, {Pringle}, \& {Naylor}}]{Dobbs+2014}
{Dobbs}, C.~L., {Pringle}, J.~E., \& {Naylor}, T. 2014, \mnras, 437, L31

\bibitem[{{Dobbs} {et~al.}(2010){Dobbs}, {Theis}, {Pringle}, \&
  {Bate}}]{Dobbs+2010}
{Dobbs}, C.~L., {Theis}, C., {Pringle}, J.~E., \& {Bate}, M.~R. 2010, \mnras,
  403, 625

\bibitem[{{D'Onghia} {et~al.}(2013){D'Onghia}, {Vogelsberger}, \&
  {Hernquist}}]{D'Onghia+2013}
{D'Onghia}, E., {Vogelsberger}, M., \& {Hernquist}, L. 2013, \apj, 766, 34

\bibitem[{{Draine} \& {Bertoldi}(1996)}]{DraineBertoldi1996}
{Draine}, B.~T., \& {Bertoldi}, F. 1996, \apj, 468, 269

\bibitem[{{Efstathiou} {et~al.}(1982){Efstathiou}, {Lake}, \&
  {Negroponte}}]{Efstathiou+1982}
{Efstathiou}, G., {Lake}, G., \& {Negroponte}, J. 1982, \mnras, 199, 1069

\bibitem[{{Egusa} {et~al.}(2009){Egusa}, {Kohno}, {Sofue}, {Nakanishi}, \&
  {Komugi}}]{Egusa+2009}
{Egusa}, F., {Kohno}, K., {Sofue}, Y., {Nakanishi}, H., \& {Komugi}, S. 2009,
  \apj, 697, 1870

\bibitem[{{Elmegreen}(1980)}]{Elmegreen1980}
{Elmegreen}, D.~M. 1980, \apj, 242, 528

\bibitem[{{Elmegreen} {et~al.}(2011){Elmegreen}, {Elmegreen}, {Yau},
  {Athanassoula}, {Bosma}, {Buta}, {Helou}, {Ho}, {Gadotti}, {Knapen},
  {Laurikainen}, {Madore}, {Masters}, {Meidt}, {Men{\'e}ndez-Delmestre},
  {Regan}, {Salo}, {Sheth}, {Zaritsky}, {Aravena}, {Skibba}, {Hinz}, {Laine},
  {Gil de Paz}, {Mu{\~n}oz-Mateos}, {Seibert}, {Mizusawa}, {Kim}, \& {Erroz
  Ferrer}}]{Elmegreen+2011}
{Elmegreen}, D.~M. {et~al.} 2011, \apj, 737, 32

\bibitem[{{Fioc} \& {Rocca-Volmerange}(1997)}]{FiocRocca-Volmerange1997}
{Fioc}, M., \& {Rocca-Volmerange}, B. 1997, \aap, 326, 950

\bibitem[{{Fujii} {et~al.}(2011){Fujii}, {Baba}, {Saitoh}, {Makino}, {Kokubo},
  \& {Wada}}]{Fujii+2011}
{Fujii}, M.~S., {Baba}, J., {Saitoh}, T.~R., {Makino}, J., {Kokubo}, E., \&
  {Wada}, K. 2011, \apj, 730, 109

\bibitem[{{Fujimoto}(1968)}]{Fujimoto1968}
{Fujimoto}, M. 1968, in IAU Symposium, Vol.~29, IAU Symposium, 453

\bibitem[{{Garcia-Burillo} {et~al.}(1993){Garcia-Burillo}, {Combes}, \&
  {Gerin}}]{Garcia-Burillo+1993}
{Garcia-Burillo}, S., {Combes}, F., \& {Gerin}, M. 1993, \aap, 274, 148

\bibitem[{{Gerritsen} \& {Icke}(1997)}]{GerritsenIcke1997}
{Gerritsen}, J.~P.~E., \& {Icke}, V. 1997, \aap, 325, 972

\bibitem[{{Gittins} \& {Clarke}(2004)}]{GittinsClarke2004}
{Gittins}, D.~M., \& {Clarke}, C.~J. 2004, \mnras, 349, 909

\bibitem[{{Glover} \& {Mac Low}(2007)}]{GloverMacLow2007a}
{Glover}, S.~C.~O., \& {Mac Low}, M.-M. 2007, \apjs, 169, 239

\bibitem[{{Gnedin} \& {Kravtsov}(2011)}]{GnedinKravtsov2011}
{Gnedin}, N.~Y., \& {Kravtsov}, A.~V. 2011, \apj, 728, 88

\bibitem[{{Gnedin} {et~al.}(2009){Gnedin}, {Tassis}, \&
  {Kravtsov}}]{Gnedin+2009}
{Gnedin}, N.~Y., {Tassis}, K., \& {Kravtsov}, A.~V. 2009, \apj, 697, 55

\bibitem[{{Goldreich} \& {Lynden-Bell}(1965)}]{GoldreichLynden-Bell1965}
{Goldreich}, P., \& {Lynden-Bell}, D. 1965, \mnras, 130, 125

\bibitem[{{Grand} {et~al.}(2015){Grand}, {Bovy}, {Kawata}, {Hunt}, {Famaey},
  {Siebert}, {Monari}, \& {Cropper}}]{Grand+2015}
{Grand}, R.~J.~J., {Bovy}, J., {Kawata}, D., {Hunt}, J.~A.~S., {Famaey}, B.,
  {Siebert}, A., {Monari}, G., \& {Cropper}, M. 2015, \mnras, 453, 1867

\bibitem[{{Grand} {et~al.}(2012{\natexlab{a}}){Grand}, {Kawata}, \&
  {Cropper}}]{Grand+2012b}
{Grand}, R.~J.~J., {Kawata}, D., \& {Cropper}, M. 2012{\natexlab{a}}, MNRAS,
  426, 167

\bibitem[{{Grand} {et~al.}(2012{\natexlab{b}}){Grand}, {Kawata}, \&
  {Cropper}}]{Grand+2012a}
---. 2012{\natexlab{b}}, MNRAS, 421, 1529

\bibitem[{{Grand} {et~al.}(2014){Grand}, {Kawata}, \& {Cropper}}]{Grand+2014}
---. 2014, \mnras, 439, 623

\bibitem[{{Grosbol}(1993)}]{Grosbol1993}
{Grosbol}, P. 1993, \pasp, 105, 651

\bibitem[{{Grosb{\o}l} \& {Dottori}(2009)}]{GrosbolDottori2009}
{Grosb{\o}l}, P., \& {Dottori}, H. 2009, \aap, 499, L21

\bibitem[{{Grosb{\o}l} {et~al.}(2004){Grosb{\o}l}, {Patsis}, \&
  {Pompei}}]{Grosbol+2004}
{Grosb{\o}l}, P., {Patsis}, P.~A., \& {Pompei}, E. 2004, \aap, 423, 849

\bibitem[{{Hernquist}(1990)}]{Hernquist1990}
{Hernquist}, L. 1990, \apj, 356, 359

\bibitem[{{Howard} \& {Byrd}(1990)}]{HowardByrd1990}
{Howard}, S., \& {Byrd}, G.~G. 1990, \aj, 99, 1798

\bibitem[{{Ishibashi} \& {Yoshii}(1984)}]{IshibashiYoshii1984}
{Ishibashi}, S., \& {Yoshii}, Y. 1984, \pasj, 36, 41

\bibitem[{{Julian} \& {Toomre}(1966)}]{JulianToomre1966}
{Julian}, W.~H., \& {Toomre}, A. 1966, \apj, 146, 810

\bibitem[{{Junqueira} {et~al.}(2013){Junqueira}, {L{\'e}pine}, {Braga}, \&
  {Barros}}]{Junqueira+2013}
{Junqueira}, T.~C., {L{\'e}pine}, J.~R.~D., {Braga}, C.~A.~S., \& {Barros},
  D.~A. 2013, \aap, 550, A91

\bibitem[{{Kalnajs}(1973)}]{Kalnajs1973}
{Kalnajs}, A.~J. 1973, Proceedings of the Astronomical Society of Australia, 2,
  174

\bibitem[{{Kawata} {et~al.}(2014){Kawata}, {Hunt}, {Grand}, {Pasetto}, \&
  {Cropper}}]{Kawata+2014}
{Kawata}, D., {Hunt}, J.~A.~S., {Grand}, R.~J.~J., {Pasetto}, S., \& {Cropper},
  M. 2014, \mnras, 443, 2757

\bibitem[{{Kim} \& {Kim}(2014)}]{KimKim2014}
{Kim}, Y., \& {Kim}, W.-T. 2014, \mnras, 440, 208

\bibitem[{{Kumamoto} \& {Noguchi}(2016)}]{KumamotoNoguchi2016}
{Kumamoto}, J., \& {Noguchi}, M. 2016, arXiv160308761

\bibitem[{{Kuno} \& {Nakai}(1997)}]{KunoNakai1997}
{Kuno}, N., \& {Nakai}, N. 1997, \pasj, 49, 279

\bibitem[{{La Vigne} {et~al.}(2006){La Vigne}, {Vogel}, \&
  {Ostriker}}]{LaVigne+2006}
{La Vigne}, M.~A., {Vogel}, S.~N., \& {Ostriker}, E.~C. 2006, \apj, 650, 818

\bibitem[{{Lee}(2014)}]{Lee2014}
{Lee}, W.-K. 2014, \apj, 792, 122

\bibitem[{{Lee} \& {Shu}(2012)}]{LeeShu2012}
{Lee}, W.-K., \& {Shu}, F.~H. 2012, \apj, 756, 45

\bibitem[{{L{\'e}pine} {et~al.}(2011){L{\'e}pine}, {Cruz}, {Scarano}, {Barros},
  {Dias}, {Pomp{\'e}ia}, {Andrievsky}, {Carraro}, \& {Famaey}}]{Lepine+2011}
{L{\'e}pine}, J.~R.~D. {et~al.} 2011, \mnras, 417, 698

\bibitem[{{Lin} \& {Shu}(1964)}]{LinShu1964}
{Lin}, C.~C., \& {Shu}, F.~H. 1964, \apj, 140, 646

\bibitem[{{Lin} \& {Shu}(1966)}]{LinShu1966}
---. 1966, Proceedings of the National Academy of Science, 55, 229

\bibitem[{{Lin} {et~al.}(1969){Lin}, {Yuan}, \& {Shu}}]{Lin+1969}
{Lin}, C.~C., {Yuan}, C., \& {Shu}, F.~H. 1969, \apj, 155, 721

\bibitem[{{Lindblad}(1963)}]{Lindblad1963}
{Lindblad}, B. 1963, Stockholms Observatoriums Annaler, 22, 5

\bibitem[{{Lord} \& {Young}(1990)}]{LordYoung1990}
{Lord}, S.~D., \& {Young}, J.~S. 1990, \apj, 356, 135

\bibitem[{{Mart{\'{\i}}nez-Garc{\'{\i}}a} \&
  {Gonz{\'a}lez-L{\'o}pezlira}(2013)}]{Martinez-GarciaGonzalez-Lopezlira2013}
{Mart{\'{\i}}nez-Garc{\'{\i}}a}, E.~E., \& {Gonz{\'a}lez-L{\'o}pezlira}, R.~A.
  2013, \apj, 765, 105

\bibitem[{{Mart{\'{\i}}nez-Garc{\'{\i}}a}
  {et~al.}(2009){Mart{\'{\i}}nez-Garc{\'{\i}}a}, {Gonz{\'a}lez-L{\'o}pezlira},
  \& {G{\'o}mez}}]{Martinez-Garcia+2009b}
{Mart{\'{\i}}nez-Garc{\'{\i}}a}, E.~E., {Gonz{\'a}lez-L{\'o}pezlira}, R.~A., \&
  {G{\'o}mez}, G.~C. 2009, \apj, 707, 1650

\bibitem[{{Mart{\'{\i}}nez-Garc{\'{\i}}a} \&
  {Puerari}(2014)}]{Martinez-GarciaPuerari2014}
{Mart{\'{\i}}nez-Garc{\'{\i}}a}, E.~E., \& {Puerari}, I. 2014, \apj, 790, 118

\bibitem[{{Michikoshi} \& {Kokubo}(2016)}]{MichikoshiKokubo2016b}
{Michikoshi}, S., \& {Kokubo}, E. 2016, arXiv1604.02987

\bibitem[{{Minchev} {et~al.}(2012){Minchev}, {Famaey}, {Quillen}, {Di Matteo},
  {Combes}, {Vlaji{\'c}}, {Erwin}, \& {Bland-Hawthorn}}]{Minchev+2012}
{Minchev}, I., {Famaey}, B., {Quillen}, A.~C., {Di Matteo}, P., {Combes}, F.,
  {Vlaji{\'c}}, M., {Erwin}, P., \& {Bland-Hawthorn}, J. 2012, \aap, 548, A126

\bibitem[{{Miyamoto} {et~al.}(2014){Miyamoto}, {Nakai}, \&
  {Kuno}}]{Miyamoto+2014}
{Miyamoto}, Y., {Nakai}, N., \& {Kuno}, N. 2014, \pasj, 66, 36

\bibitem[{{Navarro} {et~al.}(1997){Navarro}, {Frenk}, \&
  {White}}]{Navarro+1997}
{Navarro}, J.~F., {Frenk}, C.~S., \& {White}, S.~D.~M. 1997, \apj, 490, 493

\bibitem[{{Oh} {et~al.}(2015){Oh}, {Kim}, \& {Lee}}]{Oh+2015}
{Oh}, S.~H., {Kim}, W.-T., \& {Lee}, H.~M. 2015, \apj, 807, 73

\bibitem[{{Oh} {et~al.}(2008){Oh}, {Kim}, {Lee}, \& {Kim}}]{Oh+2008}
{Oh}, S.~H., {Kim}, W.-T., {Lee}, H.~M., \& {Kim}, J. 2008, \apj, 683, 94

\bibitem[{{Patsis} {et~al.}(1991){Patsis}, {Contopoulos}, \&
  {Grosbol}}]{Patsis+1991}
{Patsis}, P.~A., {Contopoulos}, G., \& {Grosbol}, P. 1991, \aap, 243, 373

\bibitem[{{Pelupessy} {et~al.}(2006){Pelupessy}, {Papadopoulos}, \& {van der
  Werf}}]{Pelupessy+2006}
{Pelupessy}, F.~I., {Papadopoulos}, P.~P., \& {van der Werf}, P. 2006, \apj,
  645, 1024

\bibitem[{{Pettitt} {et~al.}(2015){Pettitt}, {Dobbs}, {Acreman}, \&
  {Bate}}]{Pettitt+2015}
{Pettitt}, A.~R., {Dobbs}, C.~L., {Acreman}, D.~M., \& {Bate}, M.~R. 2015,
  \mnras, 449, 3911

\bibitem[{{Pettitt} {et~al.}(2016){Pettitt}, {Tasker}, \&
  {Wadsley}}]{Pettitt+2016}
{Pettitt}, A.~R., {Tasker}, E.~J., \& {Wadsley}, J.~W. 2016, arXiv1603.07801

\bibitem[{{Pichardo} {et~al.}(2003){Pichardo}, {Martos}, {Moreno}, \&
  {Espresate}}]{Pichardo+2003}
{Pichardo}, B., {Martos}, M., {Moreno}, E., \& {Espresate}, J. 2003, \apj, 582,
  230

\bibitem[{{Rix} \& {Zaritsky}(1995)}]{RixZaritsky1995}
{Rix}, H.-W., \& {Zaritsky}, D. 1995, \apj, 447, 82

\bibitem[{{Roberts}(1969)}]{Roberts1969}
{Roberts}, W.~W. 1969, \apj, 158, 123

\bibitem[{{Roberts} \& {Hausman}(1984)}]{RobertsHausman1984}
{Roberts}, Jr., W.~W., \& {Hausman}, M.~A. 1984, \apj, 277, 744

\bibitem[{{Roca-F{\`a}brega} {et~al.}(2013){Roca-F{\`a}brega}, {Valenzuela},
  {Figueras}, {Romero-G{\'o}mez}, {Vel{\'a}zquez}, {Antoja}, \&
  {Pichardo}}]{Roca-Fabrega+2013}
{Roca-F{\`a}brega}, S., {Valenzuela}, O., {Figueras}, F., {Romero-G{\'o}mez},
  M., {Vel{\'a}zquez}, H., {Antoja}, T., \& {Pichardo}, B. 2013, MNRAS, 432,
  2878

\bibitem[{{Rots}(1975)}]{Rots1975}
{Rots}, A.~H. 1975, \aap, 45, 43

\bibitem[{{Rots} {et~al.}(1990){Rots}, {Bosma}, {van der Hulst},
  {Athanassoula}, \& {Crane}}]{Rots+1990}
{Rots}, A.~H., {Bosma}, A., {van der Hulst}, J.~M., {Athanassoula}, E., \&
  {Crane}, P.~C. 1990, \aj, 100, 387

\bibitem[{{Rots} \& {Shane}(1975)}]{RotsShane1975}
{Rots}, A.~H., \& {Shane}, W.~W. 1975, \aap, 45, 25

\bibitem[{{Ro{\v s}kar} {et~al.}(2012){Ro{\v s}kar}, {Debattista}, {Quinn}, \&
  {Wadsley}}]{Roskar+2012}
{Ro{\v s}kar}, R., {Debattista}, V.~P., {Quinn}, T.~R., \& {Wadsley}, J. 2012,
  \mnras, 426, 2089

\bibitem[{{Rydbeck} {et~al.}(1985){Rydbeck}, {Hjalmarson}, \&
  {Rydbeck}}]{Rydbeck+1985}
{Rydbeck}, G., {Hjalmarson}, A., \& {Rydbeck}, O.~E.~H. 1985, \aap, 144, 282

\bibitem[{{Saitoh} {et~al.}(2008){Saitoh}, {Daisaka}, {Kokubo}, {Makino},
  {Okamoto}, {Tomisaka}, {Wada}, \& {Yoshida}}]{Saitoh+2008}
{Saitoh}, T.~R., {Daisaka}, H., {Kokubo}, E., {Makino}, J., {Okamoto}, T.,
  {Tomisaka}, K., {Wada}, K., \& {Yoshida}, N. 2008, \pasj, 60, 667

\bibitem[{{Saitoh} \& {Makino}(2009)}]{SaitohMakino2009}
{Saitoh}, T.~R., \& {Makino}, J. 2009, \apjl, 697, L99

\bibitem[{{Saitoh} \& {Makino}(2010)}]{SaitohMakino2010}
---. 2010, \pasj, 62, 301

\bibitem[{{Salo} \& {Laurikainen}(2000{\natexlab{a}})}]{SaloLaurikainen2000a}
{Salo}, H., \& {Laurikainen}, E. 2000{\natexlab{a}}, \mnras, 319, 377

\bibitem[{{Salo} \& {Laurikainen}(2000{\natexlab{b}})}]{SaloLaurikainen2000b}
---. 2000{\natexlab{b}}, \mnras, 319, 393

\bibitem[{{Sawa}(1977)}]{Sawa1977}
{Sawa}, T. 1977, \pasj, 29, 781

\bibitem[{{Scarano} \& {L{\'e}pine}(2013)}]{ScaranoLepine2013}
{Scarano}, S., \& {L{\'e}pine}, J.~R.~D. 2013, \mnras, 428, 625

\bibitem[{{Seigar} \& {James}(2002)}]{SeigarJames2002}
{Seigar}, M.~S., \& {James}, P.~A. 2002, \mnras, 337, 1113

\bibitem[{{Sellwood}(2011)}]{Sellwood2011}
{Sellwood}, J.~A. 2011, MNRAS, 410, 1637

\bibitem[{{Sellwood} \& {Binney}(2002)}]{SellwoodBinney2002}
{Sellwood}, J.~A., \& {Binney}, J.~J. 2002, MNRAS, 336, 785

\bibitem[{{Sellwood} \& {Carlberg}(1984)}]{SellwoodCarlberg1984}
{Sellwood}, J.~A., \& {Carlberg}, R.~G. 1984, ApJ, 282, 61

\bibitem[{{Sellwood} \& {Carlberg}(2014)}]{SellwoodCarlberg2014}
---. 2014, \apj, 785, 137

\bibitem[{{Shetty} {et~al.}(2007){Shetty}, {Vogel}, {Ostriker}, \&
  {Teuben}}]{Shetty+2007}
{Shetty}, R., {Vogel}, S.~N., {Ostriker}, E.~C., \& {Teuben}, P.~J. 2007, \apj,
  665, 1138

\bibitem[{{Shu} {et~al.}(1972){Shu}, {Milione}, {Gebel}, {Yuan}, {Goldsmith},
  \& {Roberts}}]{Shu+1972}
{Shu}, F.~H., {Milione}, V., {Gebel}, W., {Yuan}, C., {Goldsmith}, D.~W., \&
  {Roberts}, W.~W. 1972, \apj, 173, 557

\bibitem[{{Shu} {et~al.}(1973){Shu}, {Milione}, \& {Roberts}}]{Shu+1973}
{Shu}, F.~H., {Milione}, V., \& {Roberts}, Jr., W.~W. 1973, \apj, 183, 819

\bibitem[{{Sundelius} {et~al.}(1987){Sundelius}, {Thomasson}, {Valtonen}, \&
  {Byrd}}]{Sundelius+1987}
{Sundelius}, B., {Thomasson}, M., {Valtonen}, M.~J., \& {Byrd}, G.~G. 1987,
  \aap, 174, 67

\bibitem[{{Tanikawa} {et~al.}(2013){Tanikawa}, {Yoshikawa}, {Nitadori}, \&
  {Okamoto}}]{Tanikawa+2013}
{Tanikawa}, A., {Yoshikawa}, K., {Nitadori}, K., \& {Okamoto}, T. 2013, New A.,
  19, 74

\bibitem[{{Thomasson} \& {Donner}(1993)}]{ThomassonDonner1993}
{Thomasson}, M., \& {Donner}, K.~J. 1993, \aap, 272, 153

\bibitem[{{Thomasson} {et~al.}(1990){Thomasson}, {Elmegreen}, {Donner}, \&
  {Sundelius}}]{Thomasson+1990}
{Thomasson}, M., {Elmegreen}, B.~G., {Donner}, K.~J., \& {Sundelius}, B. 1990,
  \apjl, 356, L9

\bibitem[{{Toomre}(1981)}]{Toomre1981}
{Toomre}, A. 1981, in Structure and Evolution of Normal Galaxies, ed. S.~M.
  {Fall} \& D.~{Lynden-Bell}, 111--136

\bibitem[{{Tully}(1974)}]{Tully1974a}
{Tully}, R.~B. 1974, \apjs, 27, 415

\bibitem[{{Visser}(1980)}]{Visser1980b}
{Visser}, H.~C.~D. 1980, \aap, 88, 159

\bibitem[{{Vogel} {et~al.}(1988){Vogel}, {Kulkarni}, \&
  {Scoville}}]{Vogel+1988}
{Vogel}, S.~N., {Kulkarni}, S.~R., \& {Scoville}, N.~Z. 1988, \nat, 334, 402

\bibitem[{{Wada}(2008)}]{Wada2008}
{Wada}, K. 2008, \apj, 675, 188

\bibitem[{{Wada} {et~al.}(2011){Wada}, {Baba}, \& {Saitoh}}]{Wada+2011}
{Wada}, K., {Baba}, J., \& {Saitoh}, T.~R. 2011, \apj, 735, 1

\bibitem[{{Wada} \& {Koda}(2004)}]{WadaKoda2004}
{Wada}, K., \& {Koda}, J. 2004, \mnras, 349, 270

\bibitem[{{Wada} {et~al.}(2009){Wada}, {Papadopoulos}, \& {Spaans}}]{Wada+2009}
{Wada}, K., {Papadopoulos}, P.~P., \& {Spaans}, M. 2009, \apj, 702, 63

\bibitem[{{Wolfire} {et~al.}(1995){Wolfire}, {Hollenbach}, {McKee}, {Tielens},
  \& {Bakes}}]{Wolfire+1995}
{Wolfire}, M.~G., {Hollenbach}, D., {McKee}, C.~F., {Tielens}, A.~G.~G.~M., \&
  {Bakes}, E.~L.~O. 1995, \apj, 443, 152

\bibitem[{{Woodward}(1975)}]{Woodward1975}
{Woodward}, P.~R. 1975, \apj, 195, 61

\bibitem[{{Zhang}(1996)}]{Zhang1996}
{Zhang}, X. 1996, \apj, 457, 125

\end{thebibliography}

\end{document}